\documentclass[
    aps,
    prx,
    twocolumn,
    showpacs,
    floatfix,
    superscriptaddress,
    10pt,
    ]{revtex4-2} 

\usepackage{graphicx} 
\usepackage{xcolor} 
\usepackage{booktabs} 

\usepackage{mathtools} 
\usepackage{amssymb}
\usepackage{physics} 

\usepackage[
    colorlinks=true,
    linkcolor=red,
    citecolor=green,
    filecolor=cyan,      
    urlcolor=magenta,
    ]{hyperref} 

\usepackage{dsfont} 
\usepackage{bm} 

\newcommand{\EE}{\mathbf{E}}

\newcommand{\RR}{\mathbf{R}}

\newcommand{\rr}{\mathbf{r}}
\newcommand{\kk}{\mathbf{k}}

\newcommand{\pp}{\bm{\wp}}
\newcommand{\ii}{\mathrm{i}} 

\begin{document}

\title{Constructing mode-resolved quantum optical models for emitters in photonic crystals}


\author{Antonio~Morales-P\'erez}
\thanks{These authors contributed equally to this work.}
\affiliation{Donostia International Physics Center, Paseo Manuel de Lardizabal 4, 20018 Donostia-San Sebastian, Spain}
\affiliation{Material and Applied Physics Department, University of the Basque Country (UPV/EHU), Donostia-San Sebasti\'an, Spain}

\author{I\~naki~Garc\'ia-Elcano}
\thanks{These authors contributed equally to this work.}
\affiliation{Institute of Fundamental Physics IFF-CSIC, Calle Serrano 113b, 28006 Madrid, Spain}
\affiliation{Department of Physics, Columbia University, New York, New York 10027, USA}

\author{Chiara Devescovi}
\affiliation{
Institute for Theoretical Physics, ETH Zurich, 8093 Z\"urich, Switzerland
}

\author{Maia~G.~Vergniory}
\affiliation{Donostia International Physics Center, Paseo Manuel de Lardizabal 4, 20018 Donostia-San Sebastian, Spain}
\affiliation{D\'epartement de Physique et Institut Quantique, Universit\'e de Sherbrooke, Sherbrooke, QC J1K 2R1 Canada}

\author{Aitzol~Garc\'ia-Etxarri}
\email{aitzolgarcia@dipc.org}
\affiliation{Donostia International Physics Center, Paseo Manuel de Lardizabal 4, 20018 Donostia-San Sebastian, Spain}
\affiliation{Material and Applied Physics Department, University of the Basque Country (UPV/EHU), Donostia-San Sebasti\'an, Spain}
\author{Alejandro~Gonz\'alez-Tudela}
\email{a.gonzalez.tudela@csic.es}
\affiliation{Quantum Advanced Research Center (QuARC), CSIC, Calle Serrano 113b, 28006 Madrid, Spain}
\affiliation{Institute of Fundamental Physics (IFF), CSIC, Calle Serrano 113b, 28006 Madrid, Spain}

\begin{abstract}
Recent advances are enabling quantum emitters to interact with photonic crystals, whose electromagnetic modes exhibit complex dispersion relations, spatial mode structure, and polarization textures. However, modeling light-matter behavior in these systems faces a persistent trade-off: electromagnetic approaches based on Maxwell-equation solvers provide realistic vectorial descriptions but are difficult to integrate with quantum many-body and non-perturbative methods, whereas simplified quantum-optical lattice models are tractable but typically rely on scalar and spatially independent light–matter couplings that miss essential features of these structured photonic environments. Here, we introduce a constructive framework to derive quantum-optical lattice descriptions that overcome this trade-off. Combining symmetry-constrained tight-binding constructions with numerically computed photonic band structures and field profiles, our method yields minimal, symmetry-enforced lattice Hamiltonians that reproduce the target photonic dispersion while retaining the mode-resolved (position- and polarization-dependent) structure of the light–matter coupling. We show that these models recover Green’s-function-based emitter dynamics in the perturbative regime, while providing access to non-perturbative quantum dynamical simulations beyond emitter-only descriptions. As a proof of principle, we apply the framework to a two-dimensional photonic crystal and show that it captures polarization-dependent directional emission inaccessible to scalar models, while enabling the analysis of non-Markovian light–matter dynamics and entanglement. Our results provide a practical bridge between classical electromagnetic simulation tools and quantum-optical many-body and non-Markovian modeling in photonic crystal settings.
\end{abstract}

\maketitle


\section{Introduction} \label{sec:intro}

Quantum emitters coupled to photonic crystals~\cite{goban13a,thompson13a,goban15a,hood16a,Beguin2020a,Zhou2023,Kim2019a,Samutpraphoot2020,Dordevic2021,Menon2024,Tiranov2023,evans18a,Lukin2022a,Rugar2020,Rugar2021,Lange2024,Hansen2025,Fayard2022a,Bouscal2024,Margalit2026} provide a versatile platform for engineering light-matter interactions through tailored electromagnetic environments~\cite{Sheremet2023WaveguideCorrelations,Chang2018,Gonzalez-Tudela2024,Dagget2026}. By shaping photonic dispersion, spatial mode structure, and polarization textures, these systems enable phenomena ranging from directional and topological photonic transport~\cite{Yu2019,ozawa19a,Rider2019,Mehrabad2023TopologicalDirections,lodahl17a,Suarez2025} to emitter-induced photon localization~\cite{bykov75a,john90a,kurizki90a,Gonzalez-Tudela2015,douglas15a}, non-Markovian dynamics~\cite{john94a,tanaka06a,garmon13a,garmon2019,calajo16a,Gonzalez-Tudela2018,Calajo2019b,Gonzalez-Tudela2019a,deBernardis2021,rocatti2022,Vega2023TopologicalQED,Garcia-Elcano2021,rocatti2024,Vicenzio2025}, collective effects~\cite{Gonzalez-Tudela2017b,Gonzalez-Tudela2017a,galve17a,Vega2021a,Garcia-Elcano2020,Bello2019a,Leonforte2020b,GarciaElcano2023,Bello2023,Dibenedetto2025,Leonforte2024,Du2024,Salinas2024,Lanuza2022,Tecer2026}, and many-body~\cite{Shi2018,Bello2022,Rocatti2025a,Tecer2024,Lanuza2024,calajo2025,Dinc2019,Capurso2026,Barahona2025,Gonzalez2026,Windt2023,Windt2025,Windt2026,Wang2020b,Misselwitz2026,Zhang2025b} quantum-optical phenomena qualitatively distinct from free-space or cavity-QED scenarios. At the same time, this richness makes their theoretical description challenging, since emitter dynamics depend on the polarization and spatial structure of the electromagnetic modes. Developing theoretical frameworks that capture this complexity while remaining computationally tractable is thus a key challenge in quantum nanophotonics.

\begin{figure*}
    \centering
    \includegraphics[width=\linewidth]{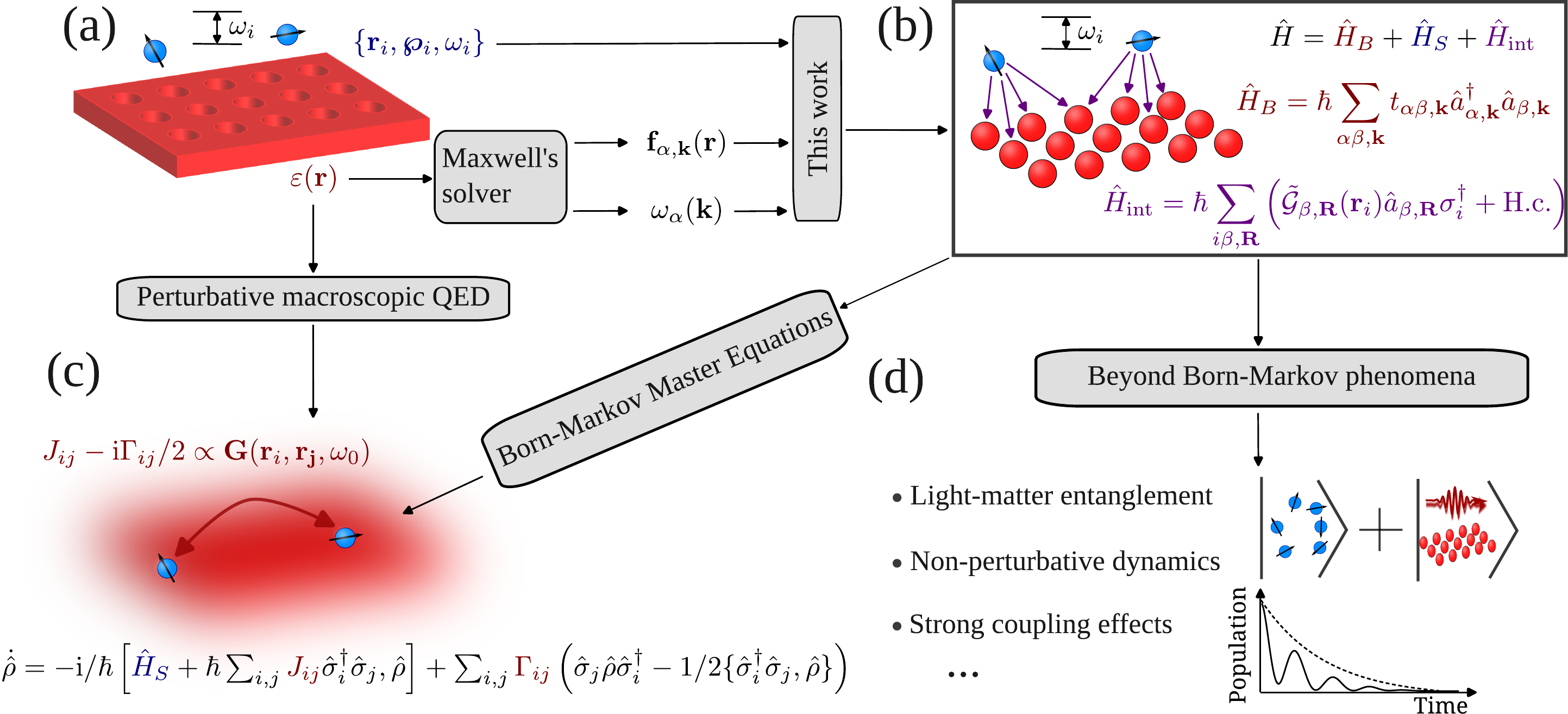}
    \caption{\textbf{Summary of the framework developed in this work.}
(a) The input data to the framework are the material's periodic dielectric function within the unit cell, $\varepsilon(\rr)$, from which we calculate the photonic band structure, $\omega_{\alpha}(\kk)$, and the corresponding electric field profiles, $\mathbf{f}_{\alpha,\kk}(\rr)$, using a Maxwell-equation solver. The emitter positions, $\rr_i$, polarization vectors, $\pp_i$, and transition frequencies, $\omega_i$, are also specified as inputs.
(b) These quantities, together with the symmetry-based methodology developed in Ref.~\cite{moralesperez2023}, are used to construct quantum-optical light-matter Hamiltonians with mode-resolved couplings described by $\hat{H} = \hat{H}_B + \hat{H}_S + \hat{H}_{\mathrm{int}}$. The photonic bath, $\hat{H}_B$, is formulated in terms of symmetry-enforced bosonic orbital operators, $\hat{a}_{\beta,\kk}$, while the interaction Hamiltonian, $\hat{H}_{\mathrm{int}}$, encodes the position- and polarization-dependent couplings, $\tilde{\mathcal{G}}_{\beta,\mathbf{R}}(\rr_i)$, between localized photonic modes, $\hat{a}_{\beta,\mathbf{R}}$ (obtained by Fourier transforming $\hat{a}_{\beta,\kk}$), and the emitter operators, $\hat{\sigma}_i^\dagger$.
(c) In Section~\ref{subsec:perturbativeconsistence}, we show that this effective model recovers the perturbative emitter dynamics predicted by macroscopic QED by adiabatically eliminating the photonic degrees of freedom, leading to an effective Born-Markov master equation for the system density matrix, $\hat{\rho}$. The resulting coherent dipole-dipole interactions, $J_{ij}$, and collective decay rates, $\Gamma_{ij}$, are expressed in terms of the electromagnetic Green's tensor, $\mathbf{G}(\rr_i, \rr_j, \omega_0)$, assuming $\omega_i\equiv\omega_0$.
(d) Beyond this regime, keeping explicitly the photonic degrees of freedom enables exact quantum dynamical simulations beyond the Born–Markov approximation, providing access to non-Markovian dynamics, strong-coupling effects, and quantum observables such as light–matter entanglement.}
    \label{fig:1}
\end{figure*}

This challenge arises because accurately capturing light-matter interactions in photonic crystals entails a fundamental trade-off between microscopic realism and quantum-optical tractability. Classical electromagnetic approaches based on Maxwell-equation solvers provide realistic descriptions of emitter interactions in structured media~\cite{Gruner1996,dung02a,buhmann07a,Gonzalez-Tudela2011,dzsotjan10a,asenjogarcia17a,Perczel2020a,Navarro-Baron2021a,blanco2022energy,wu2014plasmon,anger2006enhancement,garcia2019field,artuso2011using,neuman2015mapping}, but are difficult to integrate with quantum many-body methods, a limitation that becomes particularly important when addressing non-perturbative regimes. Spectral-density fitting approaches~\cite{Feist2020,Medina2021,SanchezBarquilla2022,SanchezMartinez2024} and quasi-normal-mode expansions~\cite{Ching1998,Kristensen2014,Franke2019} can be embedded into non-perturbative quantum descriptions, but typically remain tied to computationally demanding electromagnetic calculations or to reduced descriptions of selected resonances and reservoir responses. Reduced frameworks have long been developed to overcome this limitation. Coupled-mode theory~\cite{Haus1984,Fan2003} and tight-binding models built from photonic Wannier functions~\cite{Albert2000,Whittaker2003,Busch2003,wolff2013generation,moralesperez2023} provide compact representations of the electromagnetic band structure. However, they focus on photonic propagation and do not incorporate the coupling to quantum emitters. Complementary quantum-optical emitter-bath models based on effective tight-binding Hamiltonians~\cite{calajo16a,Gonzalez-Tudela2018,Calajo2019b,Gonzalez-Tudela2019a,deBernardis2021,rocatti2022,Vega2023TopologicalQED,Garcia-Elcano2021,rocatti2024,Vicenzio2025,Gonzalez-Tudela2017b,Gonzalez-Tudela2017a,galve17a,Vega2021a,Garcia-Elcano2020,Bello2019a,Leonforte2020b,GarciaElcano2023,Bello2023,Dibenedetto2025,Leonforte2024,Du2024,Salinas2024,Lanuza2022,Tecer2026,Shi2018,Bello2022,Rocatti2025a,Tecer2024,Lanuza2024,calajo2025,Dinc2019,Capurso2026,Barahona2025,Gonzalez2026,Windt2023,Windt2025,Windt2026,Wang2020b,Misselwitz2026,Zhang2025b} are well suited to analytical tools and many-body numerical techniques such as resolvent methods~\cite{cohentannoudji98a} and tensor networks~\cite{Orus2019}. However, they often rely on scalar or phenomenological light-matter couplings that neglect the emitter's position and dipole-orientation dependence inherited from the electromagnetic mode profiles, which is essential for phenomena such as chiral quantum optics~\cite{ozawa19a,Rider2019,Mehrabad2023TopologicalDirections,lodahl17a,Suarez2025}. Therefore, constructing computationally tractable quantum-optical models that retain realistic position- and polarization-dependent light-matter couplings in photonic crystals remains an open challenge.

In this work, we address this challenge by introducing a constructive framework to derive minimal quantum-optical lattice models with mode-resolved light-matter couplings for emitters coupled to photonic crystals. Starting from electromagnetic data obtained using Maxwell-equation solvers~\cite{johnson01a} and the emitter positions, see Fig.~\ref{fig:1}, our approach combines symmetry-constrained tight-binding constructions based on elementary band representations~\cite{Christensen2023,moralesperez2023} with numerically computed photonic band structures and field profiles to build lattice Hamiltonians that capture the target photonic dispersion, including its symmetry properties and topological character, while retaining the position- and polarization-dependent structure of the light–matter coupling. In the perturbative regime, we show that these models recover the predictions of Green’s-function-based emitter dynamics, while also providing access to non-perturbative quantum dynamical regimes beyond emitter-only descriptions. As an illustrative example, we apply the framework to a two-dimensional photonic crystal and show that it captures polarization-dependent directional emission~\cite{Yu2019} inaccessible to scalar reduced models~\cite{Gonzalez-Tudela2017a,Gonzalez-Tudela2017b}, while enabling the study of quantum dynamical phenomena such as non-Markovian light–matter dynamics and entanglement.

The paper is organized as follows. In Section~\ref{sec:realistic}, we introduce our constructive framework for deriving minimal quantum-optical lattice models with mode-resolved light–matter couplings, and demonstrate that the resulting models recover the effective emitter dynamics predicted by macroscopic QED in the perturbative regime. In Section~\ref{sec:example}, we present a proof-of-principle application to a two-dimensional photonic crystal, demonstrating the ability of the approach to capture polarization-dependent directional emission and non-Markovian quantum dynamics. Finally, in Section~\ref{sec:conclusions}, we summarize our results and outline future research directions.

\section{Constructive framework for minimal quantum-optical lattice models} \label{sec:realistic}

In this section, we present the constructive framework used to derive minimal quantum-optical lattice models for $N_e$ emitters coupled to photonic crystals. These models are described by a light-matter Hamiltonian consisting of three parts, $\hat{H} = \hat{H}_B + \hat{H}_S + \hat{H}_{\mathrm{int}}$, corresponding to the photonic bath, the emitters, and their light-matter coupling, respectively. The central idea is to project the structured electromagnetic description onto a symmetry-constrained reduced basis while preserving the position and polarization dependence of light-matter couplings. To do so, in Section~\ref{subsec:effectivebath} we first construct an effective tight-binding Hamiltonian $\hat{H}_B$ for the photonic subsystem in terms of symmetry-enforced orbitals~\cite{Christensen2023,moralesperez2023}. In Section~\ref{subsec:effectivecoupling}, we show how to reconstruct the corresponding electric field and derive the effective light-matter interaction Hamiltonian $\hat{H}_\mathrm{int}$ in this reduced basis. Finally, in Section~\ref{subsec:perturbativeconsistence}, we demonstrate that, in the perturbative regime, this framework recovers the effective Born-Markov emitter dynamics obtained from the macroscopic QED formalism.

\subsection{Minimal photonic tight-binding model} \label{subsec:effectivebath}

The first step of our framework is to replace the full electromagnetic description of the photonic crystal by a reduced lattice bosonic model that faithfully reproduces the $N_b$ photonic bands $\omega_\alpha(\kk)$ relevant for the emitter dynamics, namely those energetically close to the emitter transition frequencies $\omega_i$. Concretely, starting from the photonic band structure obtained from a Maxwell-equation solver, we seek an effective bosonic lattice Hamiltonian of the form:
\begin{equation} \label{eq:HB}
    \hat{H}_B = 
    \sum_{\kk} 
    \hat{\mathbf{A}}_\kk^\dagger \mathbf{H}(\kk) \hat{\mathbf{A}}_\kk = \hbar \sum_{\kk,\alpha,\beta} t_{\alpha,\beta,\kk}\hat{a}^\dagger_{\alpha,\kk}\hat{a}_{\beta,\kk},
\end{equation}
where $\hat{\mathbf{A}}_\kk= \left[ \hat{a}_{1,\kk}, \dots, \hat{a}_{N_o,\kk}\right]$ collects effective photonic annihilation operators, and $\mathbf{H}(\kk)$ is a $N_o\times N_o$ Hermitian matrix with matrix elements $t_{\alpha,\beta,\kk}$, whose eigenvalues reproduce the target photonic dispersion. Here, $N_o$ denotes the number of basis functions (orbitals) required for the model to reproduce $\omega_\alpha(\kk)$.

In principle, one could fit the target photonic bands using a generic tight-binding Hamiltonian, in which case the required number of orbitals, $N_o$, depends on both the fitting ansatz and the desired accuracy. Without additional constraints, however, such constructions are largely arbitrary and provide no systematic route toward minimal, physically meaningful reduced models, where minimality refers to the smallest possible orbital number $N_o$. To remove this ambiguity, we employ the symmetry-based construction rooted in band representation theory introduced in Refs.~\cite{Christensen2023,moralesperez2023,hwang2026buildingblockstopologicalband}, which identifies the minimal set of localized orbitals compatible with the symmetry properties of the target photonic bands. Consequently, the number $N_o$, spatial locations, and symmetry characters of the effective orbitals $\hat{a}_{\beta,\kk}$ are completely determined by the symmetry representation content of the target bands. For effectively one- and two-dimensional systems, where Maxwell's equations decompose into scalar sectors, this construction always yields $N_o=N_b$. In three dimensions, by contrast, additional orbitals are generally required to satisfy the transversality condition of Maxwell's equations~\cite{Christensen2023,moralesperez2023,devescovi2024axion,hwang2026buildingblockstopologicalband}, see Appendix~\ref{app:TQCnew} for a brief discussion of this case.

The procedure for obtaining Eq.~\eqref{eq:HB} is as follows (a more complete discussion is given in Appendix~\ref{app:TQCnew} and Ref.~\cite{moralesperez2023}). Starting from the dielectric profile $\varepsilon(\rr)$ of the photonic crystal, we first solve the Maxwell eigenvalue problem:
\begin{equation}
\hat{\Theta} \,\mathbf{f}_{n,\kk}(\rr)=
\omega^2_{n}(\kk)\, \mathbf{f}_{n,\kk}(\rr),\label{eq:Max}
\end{equation}
with $\hat{\Theta}=\frac{c^2}{\varepsilon(\rr)}\nabla \times \nabla \times$ being the Maxwell operator~\cite{Joannopoulos2011}. This yields the complete photonic band structure together with the corresponding Bloch modes $\mathbf{f}_{n,\kk}(\rr)$, from which we select the subset of target bands relevant for the emitter dynamics.

At high-symmetry points $\kk_\nu$ of the Brillouin zone, the Bloch modes transform according to irreducible representations associated with the crystal symmetries, allowing the relevant bands to be characterized by their symmetry representation content. This information is summarized by the symmetry vector:
\begin{equation}
\bm{\eta} = \left[c_{\Gamma_1}\Gamma_1,\dots,c_{\kk_\nu}\kk_{\nu},\dots \right],
\end{equation}
where $c_{\kk_\nu}$ denotes the multiplicity of the corresponding irreducible representation at the high-symmetry point $\kk_\nu$~\footnote{We adopt the notation of the Bilbao Crystallographic Server~\cite{aroyo2011crystallography}, where the irreps and elementary band representations are tabulated.}. The next step is therefore to determine the symmetry vector $\bm{\eta}$ associated with the target bands from their symmetry properties at the relevant high-symmetry points.

When $\bm{\eta}$ can be expressed as a linear combination of elementary band representations with non-negative integer coefficients~\cite{Christensen2023,moralesperez2023}, the target bands admit a description in terms of symmetry-adapted orbitals $\{\ket{\phi_{\mu,i}(\rr)}\}$ localized at Wyckoff positions $\boldsymbol{\delta}_{\mu}$. Here, $i$ labels the dimension of the corresponding irreducible representation and $\mu$ labels the different Wyckoff positions. Fourier transforming these localized orbitals yields a symmetry-adapted Bloch basis, $\mathcal{F}[\ket{\phi_{\mu,i}(\rr)}]=\ket{\psi_{\mu,i}(\kk)}$, in which the Maxwell operator is represented by the matrix
\begin{equation}
[\bm{\Theta}(\kk)]_{\nu,i,\mu,j}
=
\bra{\psi_{\nu,i}(\kk)}
\hat{\Theta}
\ket{\psi_{\mu,j}(\kk)}\,.
\label{eq:Maxwell}
\end{equation}

The chosen orbital content fixes the symmetry-allowed structure of the matrix $\bm{\Theta}(\kk)$, while the truncation of the hopping range controls the complexity and accuracy of the reduced model. Together, these constraints leave only a finite set of free parameters, which are determined by fitting the eigenvalues $\lambda_\alpha(\kk)$ of $\bm{\Theta}(\kk)$ to the squared target photonic dispersion, i.e., $
\lambda_\alpha(\kk)\approx\omega_\alpha^2(\kk)$.

Finally, we promote the symmetry-adapted Bloch basis to bosonic degrees of freedom by associating creation and annihilation operators $\hat{a}^{\dagger}_{\beta,\kk}$ and $\hat{a}_{\beta,\kk}$ with each Fourier-transformed orbital $\ket{\psi_{\mu,i}(\kk)}$. Here, we introduce the composite index $\beta=\{\mu,i\}$, which combines the Wyckoff-position label $\mu$ and the irreducible-representation index $i$. The Bloch Hamiltonian of Eq.~\eqref{eq:HB} is then obtained from the projected Maxwell operator through
\begin{equation}
\mathbf{H}(\kk)=\hbar\sqrt{\bm{\Theta}(\kk)},
\end{equation}
where the square root is understood in the operator sense. After this procedure, one obtains a minimal symmetry-enforced tight-binding Hamiltonian $\hat{H}_B$ that reproduces the target photonic dispersion using only a small number of physically meaningful parameters constrained by symmetry.

\subsection{Light-matter coupling in the orbital basis}
\label{subsec:effectivecoupling}

A key advantage of the present framework over scalar reduced models is that the effective photonic description retains the microscopic information required to reconstruct mode-resolved light–matter couplings, that is, including their dependence on both the position and dipole orientation of the emitters. We model the emitters as two-level systems characterized by their transition frequencies $\omega_i$, dipole moments $\pp_i$, and positions $\rr_i$. Their Hamiltonian reads
\begin{align}
\hat{H}_S = \hbar \sum_{i=1}^{N_e} \omega_i \hat{\sigma}_i^\dagger \hat{\sigma}_i\,,
\label{eq:HS}
\end{align}
where $\hat{\sigma}_i=\ket{g}_i\bra{e}$ and $\hat{\sigma}^\dagger_i=\ket{e}_i\bra{g}$ are the lowering and raising operators associated with the optical transition of the $i$-th emitter, and $N_e$ is the total number of emitters.

Since the electric field operator is naturally expressed in terms of the physical photonic eigenmodes obtained from the Maxwell problem, we first diagonalize the reduced photonic Hamiltonian of Eq.~\eqref{eq:HB}, defining the corresponding eigen-operators as
\begin{equation}
    \hat{d}_{\alpha,\kk} =
    \sum_{\beta=1}^{N_o}
    U_{\alpha,\beta}(\kk)
    \hat{a}_{\beta,\kk},
    \label{eq:change1}
\end{equation}
where $U_{\alpha,\beta}(\kk)$ are the matrix elements of the unitary transformation that diagonalizes $\mathbf{H}(\kk)$. Using these operators together with the mode profiles $\mathbf{f}_{\alpha,\kk}(\rr)$ obtained from the Maxwell-equation solver, the electric field operator takes the form
\begin{equation}
\label{eq:Eorbital}
    \hat{\EE}(\rr) =
    \sum_{\alpha,\kk}
    \left[
    \ii\,
    C_{\alpha,\kk}\mathbf{f}_{\alpha,\kk}(\rr)
    \hat{d}_{\alpha,\kk}
    + \mathrm{H.c.}
    \right],
\end{equation}
where $C_{\alpha,\kk}$ is a normalization factor ensuring the correct physical units of the electric field operator. Its explicit form depends on the normalization convention adopted for the eigenmodes $\mathbf{f}_{\alpha,\kk}(\rr)$ and is given in Appendix~\ref{app:coupling}.

With this electric-field representation, the light-matter interaction Hamiltonian under the local dipole approximation reads
\begin{align}
\hat{H}_{\mathrm{int}}=-\sum_{i=1}^{N_{e}}\left[\pp_i \hat{\sigma}_i +\mathrm{H.c.}\right]\cdot \hat{\EE}(\rr_i)\,,~\label{eq:Hintexact}
\end{align}
which defines the full dipole light-matter Hamiltonian before making the rotating-wave approximation. In the optical regime, however, the light-matter coupling strengths are generally much smaller than the emitter and photonic frequencies. It is therefore customary to neglect the counter-rotating terms in Eq.~\eqref{eq:Hintexact}, yielding the rotating-wave Hamiltonian:
\begin{align} 
\hat{H}_{\mathrm{int}}= \hbar \sum_{i=1}^{N_{e}}\sum_{\alpha=1}^{N_o} \sum_{\kk}\left[g_{\alpha,\kk}(\rr_i)\hat{d}_{\alpha,\kk}\hat{\sigma}_i^\dagger+\mathrm{H.c.}\right]\,,
\label{eq:Hint1}
\end{align}
where the coupling coefficient is
\begin{equation}
\label{eq:couplingeig}
g_{\alpha,\kk}(\rr_i)= -\frac{\ii}{\hbar} C_{\alpha,\kk}\,\pp_i^*\cdot\mathbf{f}_{\alpha,\kk}(\rr_i)\,.
\end{equation}

Using Eq.~\eqref{eq:change1}, we can express this light-matter coupling in terms of the momentum-dependent orbitals:
\begin{align}
\hat{H}_{\mathrm{int}} = \hbar \sum_{i=1}^{N_{e}}\sum_{\beta=1}^{N_o} \sum_{\kk}\left[\mathcal{G}_{\beta,\kk}(\rr_i)\hat{a}_{\beta,\kk}\hat{\sigma}_i^\dagger+\mathrm{H.c.}\right]\,,\label{eq:Hint2}
\end{align}
with:
\begin{equation}
  \mathcal{G}_{\beta,\kk}(\rr_i) = 
    -\frac{\ii}{\hbar}  \sum_{\alpha=1}^{N_o} C_{\alpha,\kk}\, \pp_i^* 
    \cdot
    \mathbf{f}_{\alpha,\kk}(\rr_i)  U_{\alpha,\beta}(\kk).\label{eq:coupling2}
\end{equation}

For numerical simulations, it is often convenient to express the interaction Hamiltonian in real space by Fourier transforming the symmetry-enforced orbital operators:
\begin{equation}
    \hat{a}_{\beta,\kk} = 
    \frac{1}{\sqrt{N}}
    \sum_\RR
    e^{-i\kk\cdot(\RR+\bm{\delta}_\beta)}
    \hat{a}_{\beta,\RR},
\end{equation}
where $\RR$ runs over lattice vectors, $N$ is the number of lattice sites, and $\bm{\delta}_\beta$ defines the Wyckoff position of orbital $\beta$. Using this definition, the light-matter Hamiltonian then reads:
\begin{align}
\hat{H}_{\mathrm{int}} = \hbar \sum_{i=1}^{N_{e}}\sum_{\beta=1}^{N_o} \sum_{\RR}\left[\tilde{\mathcal{G}}_{\beta,\RR}(\rr_i)\hat{a}_{\beta,\RR}\hat{\sigma}_i^\dagger+\mathrm{H.c.}\right]\,,\label{eq:Hint3}
\end{align}
where
\begin{equation}
    \tilde{\mathcal{G}}_{\beta,\RR}(\rr_i) = 
    \frac{1}{\sqrt{N}}
    \sum_\kk
    \mathcal{G}_{\beta,\kk}(\rr_i)
    e^{-i\kk\cdot(\RR+\bm{\delta}_\beta)}\,,\label{eq:coupling3}
\end{equation}
representing the coupling between the emitter at position $\rr_i$ and the $\beta$-th orbital located at lattice site $\RR$.  Let us emphasize that, unlike scalar reduced models, the couplings derived in this section retain the full dependence on emitter position and dipole orientation inherited from the underlying electromagnetic mode structure.

\subsection{Perturbative consistency with macroscopic QED}
\label{subsec:perturbativeconsistence}

To validate the minimal light-matter model, we now show that, in the perturbative regime, it reproduces the effective emitter dynamics obtained within macroscopic QED in the same limit~\cite{Gruner1996,dung02a,buhmann07a,Gonzalez-Tudela2011,dzsotjan10a,asenjogarcia17a,Yu2019,Perczel2020a,Navarro-Baron2021a}. Throughout this section, for the purpose of the perturbative comparison, we retain the counter-rotating terms of Eq.~\eqref{eq:Hintexact} so that the resulting perturbative coefficients coincide with those obtained from the two-pole Green's tensor in macroscopic QED. The complete derivation is presented in Appendix~\ref{app:effect}.

In macroscopic QED, the electric field operator is expressed in terms of the classical Green's tensor, $\mathbf{G}(\rr, \rr', \omega)$. In the perturbative regime, where the photonic degrees of freedom can be adiabatically eliminated, the emitter dynamics are governed by the Born-Markov master equation:
\begin{align} 
    \dot{\hat{\rho}} &= 
    -\frac{\ii}{\hbar} 
    \left[ \hat{H}_S + 
    \hbar\sum_{i\neq j}^{N_e} J_{ij} \hat{\sigma}_i^\dagger \hat{\sigma}_j, \hat{\rho} \right] \nonumber \\
    &+ \sum_{i,j=1}^{N_e} 
    \Gamma_{ij} 
    \left( 
      \hat{\sigma}_j \hat{\rho} \hat{\sigma}_i^\dagger 
    - \frac{1}{2} \{ \hat{\sigma}_i^\dagger \hat{\sigma}_j, \hat{\rho} \} 
    \right),\label{eq:BMa}
\end{align}
where $\hat{\rho}$ is the reduced density matrix describing the emitters' state. The coherent and dissipative photon-mediated interactions are described by $J_{ij}$ and $\Gamma_{ij}$, respectively, which are given in terms of the Green's function as:
\begin{equation}
J_{ij}(\omega_0) - \ii \frac{\Gamma_{ij}(\omega_0) }{2} = - \frac{\omega_0^2}{\hbar \varepsilon_0 c^2}
\pp_i^* \cdot \mathbf{G}(\rr_i, \rr_j, \omega_0) \cdot \pp_j\,.\label{eq:coeffG}
\end{equation}

For simplicity, this expression assumes identical emitters, $\omega_i\equiv \omega_0$, although it remains a good approximation when the emitter frequencies vary over a range small compared to the spectral scale on which $\mathbf{G}$ changes~\cite{migueltorcal2025}.

For lossless periodic media, the Green's tensor admits a spectral decomposition in terms of Bloch eigenmodes $\mathbf{f}_{n,\kk}(\rr)$~\cite{Perczel2020a}:
\begin{align} \label{eq:GFspectral1}
    \mathbf{G}(\rr, \rr', \omega_0) &= 
    \frac{c^2}{N a^3} \sum_{n} 
    \sum_\kk
    \frac
    {\mathbf{f}_{n,\kk}(\rr) \otimes \mathbf{f}_{n,\kk}^*(\rr')}
    { \omega_{n}^2(\kk) -(\omega_0 + \ii 0^+)^2}\nonumber \\
    &\approx  \frac{c^2}{2N \omega_0 a^3} \sum_{\alpha} 
    \sum_\kk
    \frac
    {\mathbf{f}_{\alpha,\kk}(\rr) \otimes \mathbf{f}_{\alpha,\kk}^*(\rr')}
    { \omega_{\alpha}(\kk) -\omega_0 - \ii 0^+}\,,
\end{align} 
where $a$ is the lattice constant and $N$ is the number of unit cells considered. Note that in the second line, we retain only the bands near $\omega_0$, since contributions from far-detuned bands are suppressed off resonance, and approximate $\omega^2_\alpha(\kk) - \omega_0^2 \simeq 2\omega_0 (\omega_\alpha(\kk) - \omega_0)$.

As we show in Appendix~\ref{app:effect}, adiabatic elimination of the photonic degrees of freedom in our minimal model within the same perturbative regime recovers the Born-Markov equation of Eq.~\eqref{eq:BMa}, with $J_{ij}$ and $\Gamma_{ij}$ now given by:
\begin{align}
{J}_{ij}(\omega_0) - \ii \frac{{\Gamma}_{ij}(\omega_0)}{2} 
&= -2\omega_0
\sum_{\alpha=1}^{N_o} \sum_{\kk} 
\frac{g_{\alpha,\kk}(\rr_i)\, g_{\alpha,\kk}^*(\rr_j)}{\omega_{\alpha}^2(\kk) - (\omega_0 + \ii 0^+)^2}
\nonumber \\
&\approx -
\sum_{\alpha=1}^{N_o} \sum_{\kk} 
\frac{g_{\alpha,\kk}(\rr_i)\, g_{\alpha,\kk}^*(\rr_j)}{\omega_{\alpha}(\kk) - \omega_0 - \ii 0^+}\,.
\label{eq:perturbativecoupl}
\end{align}

The first line retains both poles and reproduces the macroscopic-QED result from Eq.~\eqref{eq:GFspectral1} within the relevant-band approximation after substituting the couplings of Eq.~\eqref{eq:couplingeig}, including the contributions from both the rotating and counter-rotating terms~\footnote{Time-reversal symmetry of the lossless dielectric is also used in the derivation.}. The second line follows from the same near-resonant approximation employed in Eq.~\eqref{eq:GFspectral1}, namely retaining only bands close to $\omega_0$ and linearizing the denominator. Equivalently, it is obtained by adiabatically eliminating the photonic degrees of freedom starting directly from the rotating-wave Hamiltonian of Eq.~\eqref{eq:Hint1}.

\section{Proof-of-principle application: polarization-dependent emission in a two-dimensional photonic crystal} \label{sec:example}

\begin{figure}
    \centering
    \includegraphics[width=\linewidth]{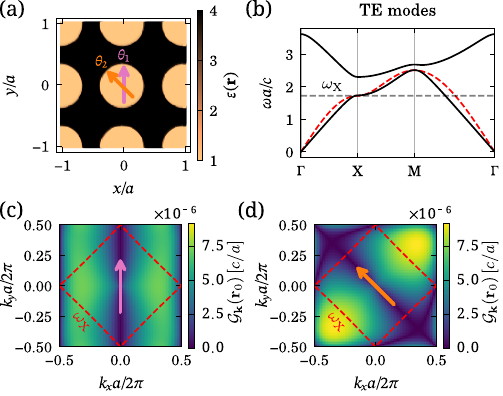}
    \caption{\textbf{Effective light-matter model for a quantum emitter coupled to a 2D photonic crystal.} (a) Dielectric permittivity profile $\varepsilon(\mathbf{r})$ within the unit cell of the photonic crystal. Magenta and orange arrows indicate the two emitter dipole orientations considered, $\theta_1=90^\circ$ and $\theta_2=135^\circ$, respectively. (b) Photonic band structure for transverse-electric modes along the high-symmetry path of the first Brillouin zone (solid black lines). The horizontal dashed line marks the saddle-point frequency $\omega_\mathrm{X}$, while the red dashed curve shows the fitted reduced-model dispersion of Eq.~\eqref{eq:tbaprox}. (c-d) Reciprocal-space coupling profile $|\mathcal{G}_{\mathbf{k}}(\mathbf{r}_0)|$ from Eq.~\eqref{eq:coupling2} over the first Brillouin zone for an emitter placed at the origin $\mathbf{r}_0$ of the unit cell, for dipole orientations (c) $\theta_1=90^\circ$ and (d) $\theta_2=135^\circ$. The magenta/orange arrows at the $\Gamma$ point indicate the dipole orientation, and the dashed red contour marks the isofrequency line $\omega_\mathbf{k}=\omega_\mathrm{X}$ obtained from the reduced-model dispersion of Eq.~\eqref{eq:tbaprox}. The coupling has appreciable weight on different sectors of the $\omega_\mathrm{X}$ contour depending on polarization. Since the group velocity is normal to the isofrequency contour, these polarization-selected $\kk$-space regions determine the real-space propagation directions of the emitted photonic excitation, leading to the qualitatively different directional emission shown in Fig.~\ref{fig:3}(c-d).}
    \label{fig:2}
\end{figure}

To illustrate the capabilities of our framework, we apply it to a quantum emitter placed near the two-dimensional photonic crystal shown in Fig.~\ref{fig:2}(a). Studying spontaneous emission in this environment provides a stringent benchmark because it exhibits strongly directional radiation patterns. As previously demonstrated for a related slab structure in Ref.~\cite{Yu2019} using semiclassical electromagnetic simulations, these patterns depend sensitively on both the emitter position within the unit cell and its dipole polarization. While simplified scalar descriptions have anticipated directional emission in structured photonic environments~\cite{Gonzalez-Tudela2017b,Gonzalez-Tudela2017a,galve17a}, they neglect the vectorial structure of the electromagnetic modes and therefore cannot reproduce the polarization-dependent effects reported in Ref.~\cite{Yu2019}. 

In this section, we show how our framework bridges this gap within a fully quantum coupled-mode description. We first derive, in Section~\ref{subsec:tbexample}, the effective light-matter model for a two-dimensional photonic crystal featuring a band structure similar to that of Ref.~\cite{Yu2019} using the methodology developed in Section~\ref{sec:realistic}. We then use this model in Section~\ref{subsec:spontaneousemission} to compute the spontaneous emission dynamics non-perturbatively for both single- and two-emitter configurations. This analysis demonstrates that our approach not only reproduces the polarization-dependent directional emission reported in Ref.~\cite{Yu2019}, but also extends beyond semiclassical perturbative treatments by capturing non-Markovian dynamics and light-matter entanglement.
 
\subsection{Constructing the mode-resolved light–matter model} \label{subsec:tbexample}

Let us now construct the mode-resolved light-matter model for the photonic structure shown in Fig.~\ref{fig:2}(a), following the procedure introduced in Section~\ref{sec:realistic}.

The structure consists of a two-dimensional square array of circular air holes with lattice constant $a$ and radius $R/a=0.355$ embedded in a dielectric medium with $\varepsilon_\mathrm{d}=4$. We first solve the Maxwell eigenvalue problem of Eq.~\eqref{eq:Max} to obtain the photonic band structure $\omega_n(\kk)$ and the corresponding mode profiles $\mathbf{f}_{n,\kk}(\rr)$. Figure~\ref{fig:2}(b) shows the first two bands, plotted as solid black lines, along a path connecting the high-symmetry points $\Gamma$, $X$, and $M$ of the Brillouin zone. The lowest band $\omega_1(\kk)$ exhibits saddle-point features accompanied by nearly straight isofrequency contours, see dashed red lines in Figs.~\ref{fig:2}(c-d). Since these spectral features were shown to be responsible for the directional emission reported in Refs.~\cite{Yu2019,Gonzalez-Tudela2017b,Gonzalez-Tudela2017a,galve17a}, we construct the effective light-matter model using only this target band $\omega_1(\kk)$. 

The next step consists of identifying the symmetries of the target band. In this case, the photonic crystal exhibits $C_2$ and $C_4$ rotational symmetries about the out-of-plane axis, together with vertical mirror symmetries, which identify the plane group as $p4mm$~\cite{aroyo2011crystallography}. Once the symmetry class is fixed, we determine the symmetry vector characterizing how the target band transforms at the high-symmetry points of the Brillouin zone. For the lowest band, this vector reads
\begin{equation}
    \bm{\eta} = \left[ \Gamma_4, \mathrm{M}_4, \mathrm{X}_2 \right].
\end{equation}

This symmetry content implies that the target band can be represented by a single localized symmetry-adapted orbital located at the origin. Therefore, the reduced model contains a single orbital degree of freedom, $N_o=1$, to which we associate a $\kk$-dependent operator $\hat{a}_{\kk}$, dropping the orbital subindex.

For this single-orbital model, the symmetry-allowed Maxwell operator contains an on-site term and hopping amplitudes between symmetry-related lattice sites. We find that keeping terms up to second-nearest neighbors is sufficient to reproduce the target band in the frequency window of interest. Equivalently, the effective photonic Hamiltonian obtained after taking the positive square root of the fitted Maxwell operator reads
\begin{equation}
\begin{split}\label{eq:tbaprox}
    \hat{H}_B & = \hbar \sum_\kk 
    \left\{ 
    J_0
    + 2 J_1 \left[ \cos(k_xa) + \cos(k_ya) \right] 
    \right.
    \\ & \hspace{1.2cm}
    \left. 
    + 4 J_2 \cos(k_xa)\cos(k_ya) 
    \right\}^{1/2}
    \hat{a}_\kk^\dagger \hat{a}_\kk,
\end{split}
\end{equation}
where the fitting parameters $J_0(a/c)^2=3.06$, $J_1/J_0=-0.26$, and $J_2/J_0=6.96\times10^{-3}$ are obtained from a least-squares fit to the squared photonic target band, constrained to reproduce the high-symmetry-point frequencies $\Gamma$, $X$, and $M$. As shown by the red dashed curve in Fig.~\ref{fig:2}(b), this three-parameter model reproduces the target band $\omega_1(\kk)$ along the high-symmetry path, including the saddle-point region that controls the directional emission benchmark.

The final step in constructing the effective light-matter model is to compute the couplings between the emitter and the reduced photonic orbital modes. To do so, we use the Maxwell eigenmodes $\mathbf{f}_{1,\kk}(\rr)$ associated with the target band $\omega_1(\kk)$ together with Eqs.~\eqref{eq:couplingeig}-\eqref{eq:coupling3}. To visualize the resulting microscopic structure, it is instructive to consider the reciprocal-space couplings $\mathcal{G}_{\kk}(\rr_0)$ between a single emitter placed at the origin of the unit cell, $\rr_0=[0,0,0]$, and the orbital modes. Figures~\ref{fig:2}(c-d) show the momentum-space coupling profile $\mathcal{G}_{\kk}(\rr_0)$ for the two representative dipole orientations, $\theta_1=90^\circ$ (magenta arrow) and $\theta_2=135^\circ$ (orange arrow), respectively, indicated in Fig.~\ref{fig:2}(a). For both polarizations, $\mathcal{G}_{\kk}(\rr_0)$ exhibits nodal lines in momentum space aligned with the dipole orientation, as well as a zero at the high-symmetry point $X$. These zeros become particularly important when they coincide with the resonant wavevectors satisfying $\omega(\kk)=\omega_0$, since they selectively suppress emission into specific propagation channels and ultimately give rise to the polarization-dependent directional emission discussed in the next section.

\subsection{Spontaneous emission dynamics and directional quantum state transfer} \label{subsec:spontaneousemission}
 
\begin{figure}
    \centering
    \includegraphics[width=\linewidth]{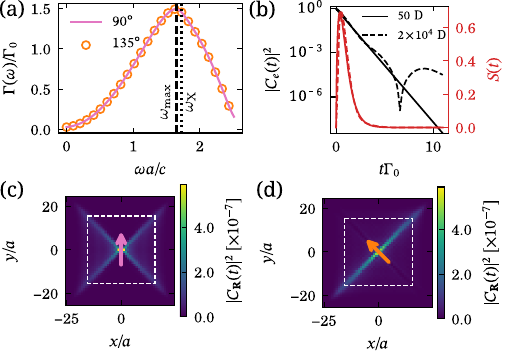}
    \caption{
\textbf{Directional spontaneous emission and non-perturbative quantum dynamics.} (a) Frequency-resolved spontaneous-emission rate $\Gamma(\omega)=\Gamma_{ii}(\omega)$ from Eq.~\eqref{eq:perturbativecoupl}, normalized to the free-space decay rate $\Gamma_0$ and evaluated with a finite broadening $0^+\rightarrow 2\pi\times10^{-2} \, c/a$, for an emitter at $\mathbf{r}_0$. Results for the two dipole orientations, $\theta_1=90^\circ$ (solid) and $\theta_2=135^\circ$ (open circles), are indistinguishable, showing that the local decay spectrum is polarization-independent at this high-symmetry position. The vertical dashed and dash-dotted lines indicate the X-point frequency $\omega_\mathrm{X}$ and the peak frequency $\omega_\mathrm{max}=\operatorname{argmax}_\omega \Gamma(\omega)$, respectively. Throughout this section, the emitter resonance is tuned to $\omega_\mathrm{max}$. (b) Time evolution of the emitter excited-state population $|C_e(t)|^2$ (black, left axis) and emitter-bath entanglement entropy $S(t)$ from Eq.~\eqref{eq:entanglement} (red, right axis), for dipole strengths $|\pp_0|=50~\mathrm{D}$ (solid) and $|\pp_0|=20000~\mathrm{D}$ (dashed), and dipole orientation $\theta_1=90^\circ$. While the smaller dipole exhibits approximately exponential decay, the larger one shows oscillations and revivals characteristic of non-Markovian dynamics accessible within the non-perturbative framework. Time is expressed in units of $\Gamma_0^{-1}$, where $\Gamma_0$ is computed self-consistently for each dipole strength. (c-d) Real-space photon population $|C_{\mathbf{R}}(t)|^2$ at time $t\Gamma_0=3 \times 10^{-2}$ for dipole orientations (c) $\theta_1=90^\circ$ and (d) $\theta_2=135^\circ$ for a dipole strength $|\pp_0|=50~\mathrm{D}$. The finite bath has $N\times N = 50\times 50$ unit cells, with the dashed white square marking the boundary of the absorbing region. All spatial coordinates are normalized by the lattice constant $a=300~\mathrm{nm}$.}
    \label{fig:3}
\end{figure}

Having constructed the effective light-matter model for this photonic crystal, we now use it to analyze the spontaneous emission dynamics beyond the perturbative regime. Our goal is twofold. First, we show that the reduced quantum-optical description reproduces the polarization-dependent directional emission expected from the full electromagnetic structure~\cite{Yu2019}, thereby validating the microscopic light-matter couplings derived in the previous section. Second, we exploit the explicit inclusion of photonic degrees of freedom in the framework to access phenomena beyond semiclassical and emitter-only treatments, including non-Markovian dynamics, bath-mediated excitation transfer, and transient entanglement generation between distant emitters.

To start, we consider the spontaneous emission of a single emitter with transition frequency $\omega_0$, dipole moment magnitude $|\pp_0|$, and position $\rr_0$ at the high-symmetry point of the unit cell. The emitter is initially prepared in its excited state, while the photonic bath is in vacuum, i.e.,
\begin{equation}
    \ket{\Psi(0)} = \ket{e} \otimes \ket{\mathrm{vac}}.
\end{equation}

As a reference point, we first study the perturbative spontaneous-emission rate $\Gamma(\omega)\equiv \Gamma_{ii}(\omega)$ predicted by the Born-Markov description introduced in Section~\ref{subsec:perturbativeconsistence}. Figure~\ref{fig:3}(a) shows the frequency dependence of $\Gamma(\omega)$, normalized to the free-space decay rate $\Gamma_0=\omega_0^3|\pp_0|^2/(3\pi\varepsilon_0\hbar c^3)$, for an emitter at $\rr_0$ and for the two dipole orientations introduced in Fig.~\ref{fig:2}. The decay spectrum exhibits a pronounced peak near the saddle-point frequency $\omega_\mathrm{X}$ due to the enhanced local density of states associated with the van Hove singularity of the photonic band structure. However, we do not observe the logarithmic divergence predicted in scalar models~\cite{Gonzalez-Tudela2017a,Gonzalez-Tudela2017b}, because the coupling $\mathcal{G}_{\kk_X}(\rr_0)=0$ at this high-symmetry emitter position $\rr_0$ independent of dipole orientation, as shown in Figs.~\ref{fig:2}(c-d). Moreover, the peak is slightly shifted to a frequency $\omega_\mathrm{max}$ relative to $\omega_\mathrm{X}$ due to the longer-range hopping terms retained in the fitted tight-binding model of Eq.~\eqref{eq:tbaprox}. In what follows, we tune the emitter frequency to $\omega_0=\omega_\mathrm{max}$ to maximize coupling to the photonic bath, unless stated otherwise.

Another noteworthy feature of Fig.~\ref{fig:3}(a) is that, at the high-symmetry position $\rr_0$, the perturbative decay rate is independent of dipole orientation. Based on this local perturbative observable alone, one would therefore expect no qualitative distinction between the two emitter polarizations. As we now show, the full quantum dynamics predicted by the effective light-matter model reveals a markedly different picture.

To compute the exact single-emitter dynamics, we note that the total Hamiltonian $\hat{H}=\hat{H}_S+\hat{H}_B+\hat{H}_\mathrm{int}$ conserves the total excitation number. Therefore, the full light-matter wavefunction remains within the single-excitation manifold and can be written at all times as
\begin{align} \label{eq:wave}
    \ket{\Psi(t)} &=
    C_e(t)\ket{e}\otimes\ket{\mathrm{vac}} \nonumber \\
    &+
    \sum_\RR
    C_{\RR}(t)\,
    \ket{g}\otimes \hat{a}^\dagger_\RR \ket{\mathrm{vac}}\,,
\end{align}
where $C_e(t)$ is the probability amplitude for finding the emitter in its excited state, while $C_{\RR}(t)$ is the amplitude for finding the emitter in its ground state and a photon occupying the orbital at lattice site $\RR$. To obtain these coefficients in Figs.~\ref{fig:3}(b-d), we solve the time-dependent Schr\"odinger equation generated by the full light-matter Hamiltonian $\hat{H}$ for a finite crystal of $50\times 50$ unit cells with lattice parameter $a=300\,\mathrm{nm}$. To suppress artificial reflections from the finite simulation boundaries, we include absorbing boundary conditions as described in Appendix~\ref{app:boundaries}.

In Figs.~\ref{fig:3}(c-d), we plot snapshots of the real-space photonic population after a time $t\Gamma_0 = 3 \times 10^{-2}$ for the two dipole orientations $\theta_1=90^\circ$ and $\theta_2=135^\circ$, respectively, using a dipole strength $|\pp_0|=50~\mathrm{D}$~\cite{gentile2016localized, guest2002measurement, li2022excitonic}. The reduced quantum-optical model reproduces the polarization-dependent directional emission previously reported in Ref.~\cite{Yu2019}, which is absent in scalar reduced descriptions~\cite{Gonzalez-Tudela2017b,Gonzalez-Tudela2017a,galve17a}. In particular, for $\theta_1=90^\circ$, the emitted excitation propagates symmetrically along all four diagonal directions, $[\pm1,\pm1]$, whereas for $\theta_2=135^\circ$, the emission is predominantly channeled along a single diagonal. This orientation dependence can be understood from the momentum-space coupling profiles $\mathcal{G}_{\kk}(\rr_0)$ shown in Fig.~\ref{fig:2}(c-d), evaluated at the resonant momenta satisfying $\omega(\kk_0)=\omega_0$, which in this case lie along four nearly straight isofrequency contours. For $\theta_1=90^\circ$, $\mathcal{G}_{\kk_0}(\rr_0)\neq 0$ at the resonant momenta $\kk_0=(\pm\pi/2,\pm\pi/2)$, which correspond to the modes with the largest group velocity. By contrast, for $\theta_2=135^\circ$, $\mathcal{G}_{\kk_0}(\rr_0)=0$ for the subset $\kk_0=(\pm\pi/2,\mp\pi/2)$, thereby suppressing one pair of propagation channels and producing the observed directional asymmetry. This picture can be quantified through the Shannon entropy of the photonic population coarse-grained over the four lattice quadrants, yielding values close to $\ln 4$ for $\theta_1$ and $\ln 2$ for $\theta_2$, consistent with four-channel and effectively bidirectional emission, respectively.

While the spatial emission pattern depends strongly on dipole orientation, the local excited-state dynamics remain nearly indistinguishable for the two polarizations at this high-symmetry position. For this reason, in Fig.~\ref{fig:3}(b) we only show the time evolution of the emitter excited-state population $|C_e(t)|^2$ for a fixed orientation $\theta_1=90^\circ$, but for two different dipole strengths. For the dipole strength $|\pp_0|=50~\mathrm{D}$, the emitter exhibits an approximately exponential decay on a timescale consistent with the perturbative rate $\Gamma(\omega_0)$ predicted in Fig.~\ref{fig:3}(a), indicating predominantly Markovian spontaneous emission. To illustrate the non-perturbative regime accessible within the same framework, we also consider an exaggerated dipole strength, $|\pp_0|=20000~\mathrm{D}$, for which the dynamics develop damped oscillations and population revivals due to the strong spectral enhancement near the saddle-point singularity~\footnote{Note that despite the large dipole moment, the maximum coupling strength $g_\mathrm{max}/\omega_0<0.1$ for which the rotating-wave approximation in $\hat{H}_\mathrm{int}$ of Eq.~\eqref{eq:Hint1} holds.}. While such non-Markovian effects were already anticipated in simplified scalar models~\cite{Gonzalez-Tudela2017b,Gonzalez-Tudela2017a}, our results show that the same physics can be captured within a mode-resolved light-matter description derived directly from the electromagnetic structure.

For completeness, we also plot in Fig.~\ref{fig:3}(b) the light-matter entanglement entropy
\begin{align} \label{eq:entanglement}
    S(t) &= -|C_e(t)|^2 \ln |C_e(t)|^2 \nonumber \\
    &-(1 - |C_e(t)|^2)\ln (1 - |C_e(t)|^2),
\end{align}
obtained from the reduced density matrix of the emitter. Since the global state remains pure and the dynamics are restricted to the single-excitation sector, this quantity is fully determined by the excited-state population $|C_e(t)|^2$, but provides a useful quantum interpretation of the spontaneous-emission process in terms of emitter-bath entanglement. Initially, the system is in a separable product state and $S(0)=0$. As the excitation becomes coherently shared between the emitter and the photonic bath, the entropy increases, reaching values close to $\ln 2$, before decreasing again as the excitation leaves the emitter subsystem.

\begin{figure}
    \centering
    \includegraphics[width=\linewidth]{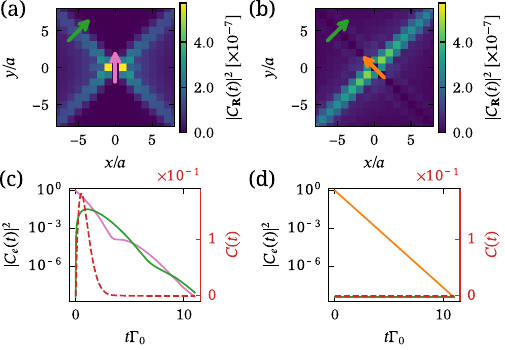}
    \caption{\textbf{Directional probe excitation and bath-mediated entanglement.} (a-b) Snapshots of the bath population $|C_{\mathbf{R}}(t)|^2$ at time $t\Gamma_0 = 6 \times 10^{-2}$ for source dipole orientations (a) $\theta_1 = 90^\circ$ and (b) $\theta_2 = 135^\circ$. The magenta/orange arrows indicate the source emitter at the origin, while the green arrow marks a probe emitter located at $\mathbf{r}_p/a = [-5,5]$ with dipole orientation $\theta_p = 45^\circ$. The finite bath has $N \times N = 50 \times 50$ unit cells, with the dashed white square marking the boundary of the absorbing region. Both emitters share the same resonance frequency $\omega_0 = \omega_\mathrm{max}$ and dipole strength $|\pp_0| = 50~\mathrm{D}$. (c-d) Time evolution for the same two source dipole orientations as in panels (a-b). The left axis (log scale) shows the source excited-state population $|C_e(t)|^2$ (magenta/orange) and probe population $|C_p(t)|^2$ (green solid), while the right axis shows the Wootters concurrence~\cite{wootters1998entanglement}, which in the present single-excitation subspace reduces to $\mathcal{C}(t)=2|C_e(t)C_p(t)|$ (red dashed), quantifying two-emitter entanglement. In panel (d), the probe population remains below the displayed range at all times and is shown clipped at the lower axis limit. Time and spatial coordinates are normalized by $\Gamma_0^{-1}$ and the lattice constant $a$, respectively.}
    \label{fig:4}
\end{figure}

The directional emission demonstrated above is not merely a single-emitter radiation effect, but can also be exploited to engineer interactions between distant quantum emitters. To illustrate this point, we now consider a two-emitter configuration in which a source emitter, initially excited and placed at the center of the photonic crystal, emits in the presence of a second probe emitter located along one of the diagonal propagation channels, at $\mathbf{r}_p/a=[-5, 5]$. As demonstrated in Fig.~\ref{fig:3}, this particular diagonal is active for one source dipole orientation but strongly suppressed for the other. The setup therefore provides a direct test of whether polarization-controlled directional emission can be used to selectively route excitations and generate controllable bath-mediated entanglement between remote emitters.

In Figs.~\ref{fig:4}(a-b), we first plot snapshots of the bath population to make the routing selectivity explicit. For the source dipole orientation $\theta_1$, the emitted photonic wavepacket propagates along the diagonal containing the probe emitter and reaches its position. In contrast, for $\theta_2$, the emission is redirected along the orthogonal diagonal and largely bypasses the probe.

This behavior is confirmed quantitatively in Figs.~\ref{fig:4}(c-d), where we plot the time evolution of the source and probe excited-state populations for the two source dipole orientations. For $\theta_1$, shown in Fig.~\ref{fig:4}(c), the probe population $|C_p(t)|^2$ develops a clear transient peak following the arrival of the wavepacket, signaling bath-mediated excitation transfer. By contrast, for $\theta_2$, shown in Fig.~\ref{fig:4}(d), the probe excitation remains strongly suppressed throughout the dynamics.

To quantify the resulting quantum correlations, we also plot the Wootters concurrence~\cite{wootters1998entanglement}, which in the present single-excitation two-emitter manifold reduces to $\mathcal{C}(t)=2|C_e(t)C_p(t)|$. For $\theta_1$, the concurrence develops a pronounced transient peak, indicating the generation of bath-mediated entanglement between the distant emitters. For $\theta_2$, the concurrence remains suppressed by several orders of magnitude. Taken together, these results demonstrate that polarization-controlled directional emission can be used to selectively switch excitation transfer and entanglement generation between remote emitters, a capability not captured by scalar or polarization-blind reduced models.

\section{Conclusions \& Outlook \label{sec:conclusions}}

Summing up, we introduce a constructive framework for deriving minimal yet realistic quantum-optical lattice models for emitters coupled to photonic crystals. Starting from first-principles electromagnetic calculations, the approach combines symmetry-constrained tight-binding constructions with the full electromagnetic mode structure to produce mode-resolved light-matter interaction Hamiltonians that retain the microscopic dependence on emitter position and dipole polarization. In the perturbative regime, the resulting effective description recovers the emitter dynamics predicted by macroscopic quantum electrodynamics, thereby establishing consistency with traditional electromagnetic approaches. Beyond this limit, the explicit inclusion of the photonic degrees of freedom enables direct access to non-perturbative quantum dynamics and unconventional radiation patterns. As a proof of principle, we apply the framework to a two-dimensional photonic crystal exhibiting saddle-point dispersion, where a minimal single-orbital model accurately captures polarization-dependent directional emission inaccessible to scalar reduced descriptions. The same framework further allows us to analyze bath-mediated entanglement generation between distant emitters in these systems.

Looking forward, the framework opens several promising directions. Methodologically, it provides a natural interface with many-body quantum-optical techniques such as tensor-network methods and resolvent-based approaches, enabling the study of collective and strongly correlated light-matter phenomena in multi-emitter systems embedded in realistic photonic crystal environments. Extensions to topological, flat-band, and multi-band photonic structures are also natural. In some of these scenarios or in strongly coupled situations, the emitter dynamics are no longer determined solely by photonic modes within a narrow spectral window around the emitter frequency, making an accurate description of the full relevant band structure increasingly important. More broadly, the explicit position- and polarization-dependent light-matter couplings derived here provide a route toward exploring quantum-optical phenomena in systems where vectorial electromagnetic effects play a central role, including chiral and topological light-matter interactions~\cite{lodahl17a, Suarez2025, ozawa19a}. We anticipate that this framework will serve as a practical bridge between electromagnetic simulation tools and quantum many-body modeling, facilitating the design of photonic-crystal platforms for quantum technologies.

\section*{Data Availability}

The numerical data and the code used to generate the photonic models and emitter dynamics reported in this article are openly available in GitHub~\cite{dataset1} and Zenodo~\cite{dataset2}.
 
\begin{acknowledgements}

AGT acknowledges support from the CSIC Research Platform on Quantum Technologies PTI-001, Spanish project Proyecto PID2024-162384NB-I00 financiado por MICIU/AEI/10.13039/501100011033 y por FEDER, UE, from the QUANTERA project MOLAR with reference PCI2024153449 and funded MICIU/AEI/10.13039/501100011033 and by the European Union, the Programa Fundamentos FBBVA through the grant EIC24-1-17304.
A.M.P, M.G.V. and A.G.E. acknowledge support from the Spanish Ministerio de Ciencia e Innovación (PID2022-142008NB-I00), from the IKUR Strategy under the collaboration agreement between Ikerbasque Foundation and DIPC on behalf of the Department of Education of the Basque Government, Programa de Ayuda de Apoyo a los agentes de la Red Vasca de Ciencia, Tecnolog\'ia e Innovaci\'on acreditados en la categor\'ia de Centros de Investigaci\'on B\'asica y de Excelencia (Programa BERC) from the Departamento de Ciencia, Universidades e Innovaci\'on del Gobierno Vasco and Centros Severo Ochoa AEI/CEX2024-0001491-S from the Spanish Ministerio de Ciencia e Innovaci\'on. A.G.E., and A.M.P also acknowledge support from the Basque Government Elkartek program (KK2025-00058). 
IGE acknowledges financial support by Programmable Quantum Materials, an Energy Frontier Research Center funded by the US Department of Energy (DOE), Office of Science, Basic Energy Sciences (BES), under award DE-SC0019443.

\end{acknowledgements}

\bibliographystyle{apsrev4-2}
\bibliography{referencesused}

\clearpage

\appendix
\onecolumngrid
\begin{center}
    \vspace{1cm}
    \textbf{\large Supplementary Information}
\end{center}

In this Supplementary Information, we provide additional technical details supporting the results presented in the main text. In Section~\ref{app:TQCnew}, we summarize the symmetry-based procedure used to construct the reduced photonic model from the full electromagnetic band structure. Section~\ref{app:coupling} details the derivation of the effective light-matter couplings and discusses the normalization conventions adopted for the electromagnetic modes. Section~\ref{app:effect} presents the derivation of the perturbative Born-Markov dynamics obtained from the reduced model and its connection to the macroscopic QED description. Section~\ref{app:boundaries} describes the numerical implementation of the real-time simulations, including the absorbing boundary conditions used to suppress finite-size reflections. Finally, Section~\ref{app:measures} defines the figures of merit used to characterize the dynamical behavior discussed in the main text.

\section{Symmetry-based construction of the reduced photonic model} \label{app:TQCnew}

In this section we summarize the procedure used to construct the symmetry-enforced reduced photonic model of Eq.~\eqref{eq:HB} in the main text. To make the construction concrete, we illustrate each step using the two-dimensional photonic crystal studied in Section~\ref{sec:example}: a square array of circular air holes with radius $R/a=0.355$ embedded in a dielectric medium with $\varepsilon_\mathrm{d}=4$. A comprehensive treatment of the general method can be found in Ref.~\cite{moralesperez2023}.

The starting point is the periodic dielectric function $\varepsilon(\rr)=\varepsilon(\rr+\RR)$ of the photonic crystal, from which one obtains the photonic eigenmodes and eigenvalues by solving the Maxwell eigenvalue problem
\begin{equation}
    \hat{\Theta}\,\mathbf{f}_{n,\kk}(\rr)
    =
    \omega_n^2(\kk)\,\mathbf{f}_{n,\kk}(\rr),
    \label{eqSM:maxwell}
\end{equation}
with the Maxwell operator
$\hat{\Theta}=\frac{c^2}{\varepsilon(\rr)}\nabla\times(\nabla\times)$.
Because $\varepsilon(\rr)$ is periodic, the eigenmodes satisfy Bloch's theorem,
$\mathbf{f}_{n,\kk}(\rr)=e^{i\kk\cdot\rr}\,\mathbf{u}_{n,\kk}(\rr)$,
where $\mathbf{u}_{n,\kk}(\rr)$ is lattice-periodic. From the full photonic spectrum, one then selects the subset of bands relevant for the emitter dynamics. For the crystal studied in Section~\ref{sec:example}, we work in the transverse-electric sector and retain only the lowest band $\omega_1(\kk)$, whose saddle-point dispersion gives rise to the directional-emission phenomena discussed in the main text [Fig.~\ref{fig:2}(b)].

At high-symmetry momenta, the Bloch modes transform according to irreducible representations (irreps) of the corresponding little groups, i.e., the subgroups of crystal symmetries that leave a given crystal momentum $\kk$ invariant up to a reciprocal lattice vector. A target set of bands is therefore characterized by a symmetry vector
\begin{equation}
    \bm{\eta} = \left[ c_{\Gamma_1}\Gamma_1,\dots,c_{\kk_\nu}\kk_{\nu},\dots \right],
\end{equation}
where $c_{\kk_\nu}$ denotes the multiplicity of irrep $\nu$ at the high-symmetry point $\kk$~\footnote{We adopt the notation established by the Bilbao Crystallographic Server~\cite{aroyo2011crystallography}.}. For the photonic crystal studied here, the $C_4$ and $C_2$ rotations about the out-of-plane axis together with the vertical mirror symmetries identify the plane group as $p4mm$~\cite{aroyo2011crystallography}. Determining the irreps of the target band at the high-symmetry points $\Gamma$, $\mathrm{X}$, and $\mathrm{M}$ yields
\begin{equation}
    \bm{\eta} = \left[ \Gamma_4, \mathrm{M}_4, \mathrm{X}_2 \right].
    \label{eqSM:eta}
\end{equation}

Throughout this work, we focus on photonic bands that can be described in terms of localized, symmetry-adapted orbitals placed at a Wyckoff position $n$ and transforming under a site-symmetry irrep $\rho$, denoted $(n|\rho)$. Each such choice induces an elementary band representation (EBR) whose symmetry content can be compared with $\bm{\eta}$. If $\bm{\eta}$ can be expressed as a linear combination of EBRs with non-negative integer coefficients, the bands form an atomic limit and are topologically trivial; if a decomposition exists only when negative integer coefficients are allowed, the bands are fragile; and if no integer combination reproduces $\bm{\eta}$, they are stably topological. Fragile bands, such as the first set of bands of most three-dimensional photonic crystals~\cite{moralesperez2023}, or topological bands can still be accommodated within this framework by adding auxiliary bands that trivialize the obstruction.
For the band of Eq.~\eqref{eqSM:eta}, the symmetry vector coincides with the single EBR induced from the $A_2$ irrep of the site-symmetry group $4mm$ at Wyckoff position $1a$, i.e. $(A_2\uparrow G)_{1a}$, hereafter $(A_2|1a)$. Because $\bm{\eta}$ matches a single EBR, the band corresponds to an atomic limit and hence is topologically trivial. Moreover, the decomposition fixes both the orbital content ($N_o=1$) and its position at the $1a$ Wyckoff position (the origin). We therefore associate a single $\kk$-dependent bosonic operator $\hat{a}_\kk$ with this orbital, dropping the orbital label.

Once a symmetry-compatible orbital basis is identified, one constructs Bloch basis functions $\psi_{\alpha,i,\kk}(\rr)$ through Fourier transformation of the localized orbitals $\{\phi_{\alpha,i}(\rr)\}$, where $\alpha$ labels symmetry-related sites within the chosen Wyckoff position and $i$ runs over the basis states of the corresponding site-symmetry irrep. Collecting these basis functions into a vector $\Psi_{\kk}(\rr)$, the action of a crystal symmetry operation $g$ is represented as
\begin{equation}
    g\,\Psi_{\kk}(\rr)=D^\top_{\kk}(g)\,\Psi_{g\kk}(\rr),
    \label{eqSM:Dk}
\end{equation}
where $D_{\kk}(g)$ is the induced representation matrix.

For the present example, the basis consists of a single $A_2$ orbital located at the $1a$ Wyckoff position. Since the origin is left invariant by the entire point group $4mm$, the induced representation is one-dimensional and $\kk$-independent, and is simply given by the characters of the $A_2$ irrep. For the generators of $p4mm$, one finds
\begin{equation}
    D_{\kk}(C_{4z})=+1,
    \qquad
    D_{\kk}(m_x)=-1,
\end{equation}
from which the remaining operations follow as consistency relations: $D_{\kk}(C_{2z})=D_{\kk}(C_{4z})^2=+1$ for the rotations, while $D_{\kk}(m_y)=-1$ and $D_{\kk}(m_{xy})=D_{\kk}(C_{4z})D_{\kk}(m_x)=-1$ for the mirror operations. In words, rotations act trivially while all mirror operations acquire a minus sign. Lattice translations act through the usual Bloch phase, $D_{\kk}(\{E|\RR\})=e^{-i\kk\cdot\RR}$.
These symmetry constraints completely determine the allowed structure of the reduced model.

In the Bloch basis, the Maxwell operator is represented by
\begin{equation}
    [\bm{\Theta}(\kk)]_{\alpha,i,\beta,j}
    =
    \bra{\psi_{\alpha,i,\kk}}
    \hat{\Theta}
    \ket{\psi_{\beta,j,\kk}},
\end{equation}
whose eigenvalues satisfy $\lambda_n(\kk)=\omega_n^2(\kk)$. For the present single-orbital example, $\bm{\Theta}(\kk)$ reduces to a scalar function $\Theta(\kk)$. To make the symmetry constraints explicit, it is convenient to first write the most general real-space hopping expansion,
\begin{equation}
    \Theta(\kk)=\sum_{\RR} J_{\RR}\,e^{i\kk\cdot\RR},
    \label{eqSM:Theta_generic}
\end{equation}
where $J_{\RR}$ denotes the hopping amplitude associated with lattice vector $\RR$. Keeping only onsite, nearest-neighbor, and next-nearest-neighbor terms introduces independent coefficients for $\RR=0$, the four nearest neighbors $\RR=\pm a\hat{x},\pm a\hat{y}$, and the four next-nearest neighbors $\RR=\pm a\hat{x}\pm a\hat{y}$.

Crystal symmetries require
\begin{equation}
    \Theta(\kk)=D_{\kk}(g)\,\Theta(g^{-1}\kk)\,D_{\kk}^\dagger(g)
    \label{eqSM:constraint}
\end{equation}
for every symmetry operation $g$. For the present single-orbital example, a major simplification occurs: since $D_{\kk}(g)=\pm1$ is a real scalar, one has $D_{\kk}(g)D_{\kk}^\dagger(g)=1$, and Eq.~\eqref{eqSM:constraint} reduces to the invariance of the dispersion under the action of the point group in momentum space,
\begin{equation}
    \Theta(\kk)=\Theta(g^{-1}\kk),
    \qquad
    g\in 4mm.
\end{equation}

Hermiticity, together with the $C_{2z}$ symmetry relating $\kk\rightarrow-\kk$, implies that all hopping amplitudes can be chosen real, i.e., $J_{-\RR}=J_\RR$. The $C_{4z}$ rotation further constrains the four nearest-neighbor hoppings to a single parameter $J_1$ and the four next-nearest-neighbor hoppings to a single parameter $J_2$, while the mirror symmetries introduce no additional constraints. Truncating Eq.~\eqref{eqSM:Theta_generic} at next-nearest neighbors, the expansion reduces to
\begin{equation}
    \Theta(\kk)=J_0+2J_1\!\left[\cos(k_xa)+\cos(k_ya)\right]+4J_2\cos(k_xa)\cos(k_ya),
    \label{eqSM:Theta_final}
\end{equation}
which is precisely the expression appearing under the square root in Eq.~\eqref{eq:tbaprox}. The remaining free parameters are then determined by fitting $\lambda(\kk)=\Theta(\kk)$ to the squared target dispersion $\omega_1^2(\kk)$.

To conclude, we promote the symmetry-adapted orbital degrees of freedom to bosonic operators by associating creation and annihilation operators $\hat{a}^\dagger_{\beta,\kk}$ and $\hat{a}_{\beta,\kk}$ with the basis functions $\psi_{\alpha,i,\kk}(\rr)$, where the composite index $\beta=\{\alpha,i\}$ combines both the label $\alpha$ of the symmetry-related sites within the Wyckoff position and the site-symmetry irrep index $i$. We then define the effective Bloch tight-binding matrix $\mathbf{H}(\kk)$ as the positive square root of the projected Maxwell matrix,
\begin{equation}
    \mathbf{H}(\kk) = \hbar \sqrt{\bm{\Theta}(\kk)},
\end{equation}
where the square root is understood in the operator sense. This step is required because the Maxwell eigenvalue problem yields eigenvalues $\omega_n^2(\kk)$, whereas the quantum Hamiltonian must be linear in the physical frequencies $\omega_n(\kk)$. For the present example, taking the square root of Eq.~\eqref{eqSM:Theta_final} reproduces the dispersion of Eq.~\eqref{eq:tbaprox}. In the present two-dimensional example this projection is trivial: the transverse-electric sector is captured by a single orbital ($N_o=N_b=1$) and $\bm{\Theta}(\kk)$ is a scalar, so no modes need to be removed before taking the square root. This is no longer the case in three dimensions, where the transversality of Maxwell's equations prevents the two lowest physical transverse bands from forming a band representation on their own; the symmetry-adapted basis must then be completed with auxiliary longitudinal orbitals, giving $N_o>N_b$~\cite{Christensen2023, moralesperez2023, devescovi2024axion, hwang2026buildingblockstopologicalband}. As shown in Ref.~\cite{moralesperez2023}, these auxiliary longitudinal orbitals sit at $\omega^2\le 0$ throughout the Brillouin zone, touching zero only at $\Gamma$, so the positive square root discards them automatically. The reduced photonic-model construction therefore carries over to three dimensions; we restrict the explicit application in this article to the two-dimensional example for simplicity.

Altogether, the Hamiltonian describing the photonic-crystal bath in the symmetry-enforced bosonic basis reads
\begin{equation}
    \hat{H}_B =
    \sum_\kk
    \hat{\mathbf{A}}_\kk^\dagger
    \mathbf{H}(\kk)
    \hat{\mathbf{A}}_\kk,
\end{equation}
where the bosonic operators are collected into the vectors $\hat{\mathbf{A}}_\kk^\dagger$ and $\hat{\mathbf{A}}_\kk$.

\section{Effective light-matter couplings and their normalization} \label{app:coupling}

In this section we derive the effective light-matter coupling constants and their normalization. Following the standard canonical quantization procedure for the electromagnetic field in transparent, dispersionless periodic dielectric media, the electric-field operator can be expanded as in Eq.~\eqref{eq:Eorbital}:
\begin{equation} \label{eq:app:Efield}
    \hat{\EE}(\rr) = \sum_{\alpha,\kk}\left[ \ii\,C_{\alpha,\kk}\,\mathbf{f}_{\alpha,\kk}(\rr)\,\hat{d}_{\alpha,\kk} + \mathrm{H.c.}\right] \,.
\end{equation}
Here, $\mathbf{f}_{\alpha,\kk}(\rr)$ are dimensionless eigenmodes of the Maxwell operator labeled by band index $\alpha$ and crystal momentum $\kk$, while $C_{\alpha,\kk}$ is a normalization coefficient with units of electric field. Substituting Eq.~\eqref{eq:app:Efield} into the dipolar interaction Hamiltonian, and invoking the rotating-wave approximation, yields the light-matter interaction term introduced in Eq.~\eqref{eq:Hint1}:
\begin{align}
    \hat{H}_{\mathrm{int}}
    = \hbar
    \sum_{i=1}^{N_e}
    \sum_{\alpha=1}^{N_o}
    \sum_{\kk}
    \left[
     g_{\alpha,\kk}(\rr_i)
     \,\hat{d}_{\alpha,\kk}\hat{\sigma}_i^\dagger
     + \mathrm{H.c.}
    \right] \, ,
\end{align}
where the coupling coefficient is given by
\begin{equation}
    g_{\alpha,\kk}(\rr_i)
    =
    -\frac{\ii}{\hbar}
    C_{\alpha,\kk}\,
    \pp_i^*\!\cdot\!\mathbf{f}_{\alpha,\kk}(\rr_i)\,.
\end{equation}

To determine the prefactor $C_{\alpha,\kk}$ explicitly, we require that the quantized electromagnetic energy reproduces the harmonic-oscillator Hamiltonian
\begin{equation}
    \hat{H}_B
    =
    \hbar\sum_{\alpha,\kk}
    \omega_{\alpha}(\kk)
    \hat{d}^{\dagger}_{\alpha,\kk}
    \hat{d}_{\alpha,\kk}
    \,.
\end{equation}
Substituting the field expansion~\eqref{eq:app:Efield}, together with the corresponding magnetic-field operator obtained from Maxwell's equation $\nabla\times\hat{\EE}=-\partial_t\hat{\mathbf{B}}$, into the electromagnetic energy functional
\begin{equation}
    \hat{H}_B
    =
    \frac{1}{2}
    \int d^3\rr
    \left[
    \varepsilon_0\varepsilon(\rr)\,\hat{\EE}^{2}(\rr)
    +
    \frac{1}{\mu_0}\hat{\mathbf{B}}^{2}(\rr)
    \right] \, ,
    \label{eq:app:HEM}
\end{equation}
one obtains spatial integrals over the total quantization volume $V_{\mathrm{tot}} = N V_{\mathrm{cell}} = Na^3$, where $N$ is the number of unit cells and $V_{\mathrm{cell}} = a^3$ is the unit-cell volume.

The Bloch eigenmodes are chosen to satisfy the orthonormality condition
\begin{equation}
    \int_{V_{\mathrm{cell}}}
    \mathbf{f}^{*}_{\alpha,\kk}(\rr)
    \cdot
    \varepsilon(\rr)
    \mathbf{f}_{\alpha',\kk}(\rr)
    \,d^3\rr
    =
    a^3\delta_{\alpha,\alpha'} \, ,
    \label{eq:app:orthonormality}
\end{equation}
which corresponds to the standard normalization convention adopted by frequency-domain Maxwell solvers such as \textsc{MPB}~\cite{johnson01a}.

Using this orthogonality relation, the electric and magnetic contributions become equal for the excitation-conserving terms, while the non-conserving contributions proportional to $\hat{d}_{\alpha,\kk}\hat{d}_{\alpha,-\kk}$ and $\hat{d}^{\dagger}_{\alpha,\kk}\hat{d}^{\dagger}_{\alpha,-\kk}$ cancel identically between the electric and magnetic sectors. The electromagnetic Hamiltonian therefore reduces to
\begin{equation}
    \hat{H}_B
    =
    \varepsilon_0 Na^3
    \sum_{\alpha,\kk}
    C_{\alpha,\kk}^{2}
    \left(
    \hat{d}^{\dagger}_{\alpha,\kk}\hat{d}_{\alpha,\kk}
    +
    \hat{d}_{\alpha,\kk}\hat{d}^{\dagger}_{\alpha,\kk}
    \right)
    \,.
    \label{eq:app:HEM_eval}
\end{equation}
Using the canonical commutation relations
$[\hat{d}_{\alpha,\kk},\hat{d}^{\dagger}_{\alpha',\kk'}]
=\delta_{\alpha,\alpha'}\delta_{\kk,\kk'}$
and discarding the vacuum-energy contribution, we obtain
\begin{equation}
    \hat{H}_B
    =
    2\varepsilon_0 Na^3
    \sum_{\alpha,\kk}
    C_{\alpha,\kk}^{2}
    \hat{d}^{\dagger}_{\alpha,\kk}\hat{d}_{\alpha,\kk}
    \,.
    \label{eq:app:HEM_diag}
\end{equation}
Matching this expression with the target harmonic-oscillator form immediately gives
\begin{equation}
    C_{\alpha,\kk}
    =
    \sqrt{\frac{\hbar\omega_{\alpha}(\kk)}{2\varepsilon_0 Na^3}}
    \,.
    \label{eq:app:Cnk}
\end{equation}
Consequently, the effective light-matter coupling takes the form
\begin{equation}
    g_{\alpha,\kk}(\rr_i)
    =
    -\ii
    \sqrt{\frac{\omega_{\alpha}(\kk)}{2\hbar\varepsilon_0 Na^3}}
    \,\pp_i^*\!\cdot\!\mathbf{f}_{\alpha,\kk}(\rr_i)
    \,.
\end{equation}
This normalization consistently relates the dimensionless classical mode profiles $\mathbf{f}_{\alpha,\kk}(\rr)$, the bosonic commutation relations of the operators $\hat{d}_{\alpha,\kk}$, and the physical dimensions of the quantized electric field.

\section{Connecting to the perturbative regime of macroscopic QED} \label{app:effect}

In this section we derive the perturbative photon-mediated interactions quoted in the main text [Eq.~\eqref{eq:perturbativecoupl}] and show that, in the relevant-band regime, our reduced light-matter model and the macroscopic-QED formalism yield the same coherent couplings $J_{ij}$ and collective decay rates $\Gamma_{ij}$ [Eq.~\eqref{eq:BMa}]. We proceed in two steps: we first obtain $J_{ij}$ and $\Gamma_{ij}$ from the reduced Hamiltonian through a standard Born-Markov elimination of the photonic bath, and then show that the same expressions follow from the spectral representation of the photonic Green's tensor.

Throughout, we retain the counter-rotating terms of the light-matter coupling, as anticipated in the main text. Besides the resonant Lamb shift and the collective decay, these terms generate an additional purely coherent (Lamb-type) shift $\Delta^c_{ij}$, associated with the anti-resonant pole of the Green's tensor, which both derivations reproduce identically.

We begin with the reduced-model derivation, using the standard Born-Markov procedure. We consider the total Hamiltonian
\begin{equation}
    \hat{H}=\hat{H}_0+\hat{H}_{\mathrm{int}},
    \qquad
    \hat{H}_0=\hat{H}_S+\hat{H}_B,
\end{equation}
with
\begin{equation}
    \hat{H}_S=\hbar\omega_0\sum_i\hat{\sigma}_i^\dagger\hat{\sigma}_i,
    \qquad
    \hat{H}_B=\hbar\sum_{\alpha,\kk}\omega_\alpha(\kk)
    \hat{d}_{\alpha,\kk}^\dagger\hat{d}_{\alpha,\kk},
\end{equation}
where, for simplicity, we assume identical emitters with $\omega_i\equiv\omega_0$. Moving to the interaction picture with respect to $\hat{H}_0$, the emitter and photonic operators evolve as
\begin{equation}
    \hat{\sigma}_i^{I}(t) = \hat{\sigma}_i\, e^{-\ii\omega_0 t},
    \qquad
    \hat{d}_{\alpha,\kk}^{I}(t) = \hat{d}_{\alpha,\kk}\, e^{-\ii\omega_\alpha(\kk) t}.
\end{equation}

Restoring the counter-rotating terms that were dropped in the rotating-wave Hamiltonian of Eq.~\eqref{eq:Hint1}, the full interaction Hamiltonian in the interaction picture reads
\begin{equation}
    \hat{H}_{\mathrm{int}}^I(t)
    =
    \hbar\sum_{i=1}^{N_e}\sum_{\alpha=1}^{N_o}\sum_{\kk}
    \left[
    g_{\alpha,\kk}(\rr_i)\hat{\sigma}_i^{\dagger,I}(t) \hat{d}_{\alpha,\kk}^{I}(t)
    +\tilde{g}_{\alpha,\kk}(\rr_i)\hat{\sigma}_i^{I}(t) \hat{d}_{\alpha,\kk}^{I}(t)
    +\mathrm{H.c.}
    \right],
    \label{eq:app:Hint_IP}
\end{equation}
where $g_{\alpha,\kk}(\rr_i)$ is the co-rotating coupling of Eq.~\eqref{eq:couplingeig}, and the counter-rotating coupling differs from it only by the absence of complex conjugation on the dipole moment,
\begin{equation}
    \tilde{g}_{\alpha,\kk}(\rr_i)
    =
    -\frac{\ii}{\hbar} C_{\alpha,\kk}\,\pp_i\!\cdot\!\mathbf{f}_{\alpha,\kk}(\rr_i)
    =
    -\ii
    \sqrt{\frac{\omega_{\alpha}(\kk)}{2\hbar\varepsilon_0 Na^3}}
    \,\pp_i\!\cdot\!\mathbf{f}_{\alpha,\kk}(\rr_i)
    \,.
    \label{eq:app:gtilde}
\end{equation}
The interaction-picture density matrix of the full light-matter system obeys the von Neumann equation
\begin{equation}
    \frac{d\hat{\rho}_{\mathrm{tot}}^I(t)}{dt}
    =
    -\frac{\ii}{\hbar}
    \left[
    \hat{H}_{\mathrm{int}}^I(t),
    \hat{\rho}_{\mathrm{tot}}^I(t)
    \right].
\end{equation}
Integrating this equation once and substituting it back yields
\begin{equation}
    \frac{d\hat{\rho}_{\mathrm{tot}}^I(t)}{dt}
    =
    -\frac{\ii}{\hbar}
    \left[
    \hat{H}_{\mathrm{int}}^I(t),
    \hat{\rho}_{\mathrm{tot}}^I(0)
    \right]
    -
    \frac{1}{\hbar^2}
    \int_0^t d\tau\,
    \left[
    \hat{H}_{\mathrm{int}}^I(t),
    \left[
    \hat{H}_{\mathrm{int}}^I(t-\tau),
    \hat{\rho}_{\mathrm{tot}}^I(t-\tau)
    \right]
    \right].
\end{equation}

We now assume that the photonic bath remains in its vacuum state and is only weakly perturbed by the emitters. Under the Born approximation, $\hat{\rho}_{\mathrm{tot}}^I(t-\tau) \approx \hat{\rho}^I(t-\tau)\otimes\ket{\mathrm{vac}}\bra{\mathrm{vac}}$, where $\hat{\rho}^I$ is the reduced emitter density matrix in the interaction picture. The first-order term vanishes after tracing over the photonic bath, since $\bra{\mathrm{vac}}\hat{d}_{\alpha,\kk}^I(t)\ket{\mathrm{vac}} = \bra{\mathrm{vac}}\hat{d}_{\alpha,\kk}^{\dagger,I}(t)\ket{\mathrm{vac}}=0$, and one obtains
\begin{equation}
    \frac{d\hat{\rho}^I(t)}{dt}
    =
    -\frac{1}{\hbar^2}
    \int_0^t d\tau\,
    \mathrm{Tr}_B
    \left[
    \hat{H}_{\mathrm{int}}^I(t),
    \left[
    \hat{H}_{\mathrm{int}}^I(t-\tau),
    \hat{\rho}^I(t-\tau)\otimes\ket{\mathrm{vac}}\bra{\mathrm{vac}}
    \right]
    \right].
\end{equation}

We next invoke the Markov approximation, replacing $\hat{\rho}^I(t-\tau)\to\hat{\rho}^I(t)$ and extending the upper integration limit to infinity. This is justified because the bath correlation function, set by the spectrum of photonic modes near $\omega_0$, decays on a timescale much shorter than that of the emitter dynamics, so the integrand contributes negligibly outside a narrow window around $\tau=0$. Using the vacuum bath correlation function:
\begin{equation}
    \bra{\mathrm{vac}}
    \hat{d}_{\alpha,\kk}^I(t)
    \hat{d}_{\beta,\kk'}^{\dagger,I}(t-\tau)
    \ket{\mathrm{vac}}
    =
    \delta_{\alpha,\beta}\delta_{\kk,\kk'}
    e^{-\ii\omega_\alpha(\kk)\tau},
\end{equation}
only pairings of a photon annihilation operator at time $t$ with a creation operator at time $t-\tau$ survive the bath trace. Two such pairings contribute: the \emph{co-rotating} one, joining the $\hat{\sigma}_i^\dagger\hat{d}_{\alpha,\kk}^{I}$ component of $\hat{H}_{\mathrm{int}}^{I}(t)$ with the $\hat{\sigma}_j\hat{d}_{\alpha,\kk}^{\dagger,I}$ component of $\hat{H}_{\mathrm{int}}^{I}(t-\tau)$, and the \emph{counter-rotating} one, joining the $\hat{\sigma}_i\hat{d}_{\alpha,\kk}^{I}$ component of $\hat{H}_{\mathrm{int}}^{I}(t)$ with the $\hat{\sigma}_j^\dagger\hat{d}_{\alpha,\kk}^{\dagger,I}$ component of $\hat{H}_{\mathrm{int}}^{I}(t-\tau)$. The remaining, mixed pairings carry an explicit $e^{\pm2\ii\omega_0 t}$ dependence on the absolute time and average to zero over the coarse-grained Markovian timescale; discarding them constitutes the secular approximation.
The two surviving emitter products are $\hat{\sigma}_i^{\dagger,I}(t)\hat{\sigma}_j^{I}(t-\tau) = \hat{\sigma}_i^\dagger\hat{\sigma}_j\, e^{\ii\omega_0\tau}$ and $\hat{\sigma}_i^{I}(t)\hat{\sigma}_j^{\dagger,I}(t-\tau) = \hat{\sigma}_i\hat{\sigma}_j^\dagger\, e^{-\ii\omega_0\tau}$. Combined with the bath factor $e^{-\ii\omega_\alpha(\kk)\tau}$, these form the single exponentials $e^{-\ii[\omega_\alpha(\kk)\mp\omega_0]\tau}$, and the reduced dynamics takes the form
\begin{align}
    \frac{d\hat{\rho}^{I}(t)}{dt}
    &=
    \sum_{i,j}\sum_{\alpha,\kk}
    g_{\alpha,\kk}(\rr_i)\,g_{\alpha,\kk}^*(\rr_j)
    \int_0^\infty\!\! d\tau\,
    e^{-\ii[\omega_\alpha(\kk)-\omega_0]\tau}
    \left[
    \hat{\sigma}_j\hat{\rho}^{I}(t)\hat{\sigma}_i^\dagger
    -
    \hat{\sigma}_i^\dagger\hat{\sigma}_j\hat{\rho}^{I}(t)
    \right]
    \nonumber\\
    &\quad+
    \sum_{i,j}\sum_{\alpha,\kk}
    \tilde{g}_{\alpha,\kk}(\rr_i)\,\tilde{g}_{\alpha,\kk}^*(\rr_j)
    \int_0^\infty\!\! d\tau\,
    e^{-\ii[\omega_\alpha(\kk)+\omega_0]\tau}
    \left[
    \hat{\sigma}_j^\dagger\hat{\rho}^{I}(t)\hat{\sigma}_i
    -
    \hat{\sigma}_i\hat{\sigma}_j^\dagger\hat{\rho}^{I}(t)
    \right]
    +\mathrm{H.c.},
    \label{eq:app:premarkov}
\end{align}
where the first line is the co-rotating channel and the second line is the counter-rotating contribution. The two memory integrals are the half-Fourier transforms of the bath correlation function evaluated at the respective detunings,
\begin{equation}
    \int_0^\infty d\tau\,
    e^{-\ii[\omega_\alpha(\kk)\mp\omega_0]\tau}
    =
    \frac{1}{\ii[\omega_\alpha(\kk)\mp\omega_0]+0^+}\,.
    \label{eq:app:tauint}
\end{equation}
For the co-rotating channel ($-\omega_0$), the real part $\pi\delta[\omega_\alpha(\kk)-\omega_0]$ is supported on the resonant modes and produces both a coherent shift and a decay; performing the mode sum gives the complex resonant kernel
\begin{equation}
    \Delta^{r}_{ij}(\omega_0)-\ii\frac{\Gamma_{ij}(\omega_0)}{2}
    = -
    \sum_{\alpha=1}^{N_o}\sum_{\kk}
    \frac{
    g_{\alpha,\kk}(\rr_i)\,
    g_{\alpha,\kk}^*(\rr_j)
    }{
    \omega_\alpha(\kk)-\omega_0 - \ii 0^+
    }.
    \label{eq:app:JGammaBM}
\end{equation}
For the counter-rotating channel ($+\omega_0$), the resonance condition $\omega_\alpha(\kk)+\omega_0=0$ is never met for physical modes, so the corresponding delta function has no support. Thus, this channel is purely dispersive and does not contribute to the decay, leading only to a dispersive coherent shift, which reads:
\begin{equation}
    \Delta^{c}_{ij}(\omega_0)
    =
    \sum_{\alpha=1}^{N_o}\sum_{\kk}
    \frac{
    \tilde{g}_{\alpha,\kk}(\rr_i)\,
    \tilde{g}_{\alpha,\kk}^*(\rr_j)
    }{
    \omega_\alpha(\kk)+\omega_0
    }.
    \label{eq:app:Deltac}
\end{equation}
Substituting Eqs.~\eqref{eq:app:JGammaBM} and \eqref{eq:app:Deltac} back into Eq.~\eqref{eq:app:premarkov}, the co-rotating channel generates the resonant coherent shift $\Delta^{r}_{ij}$ together with the collective dissipator $\Gamma_{ij}$, while the counter-rotating channel contributes the purely coherent term $+\ii\sum_{i,j}\Delta^{c}_{ij}[\hat{\sigma}_i\hat{\sigma}_j^\dagger,\hat{\rho}^I]$. Transforming back to the Schr\"odinger picture, the reduced dynamics retains the Lindblad form
\begin{align}
    \dot{\hat{\rho}}
    &=
    -\frac{\ii}{\hbar}
    \left[
    \hat{H}_S+\hat{H}_{\mathrm{eff}},
    \hat{\rho}
    \right]
    +
    \sum_{i,j}\Gamma_{ij}(\omega_0)
    \left(
    \hat{\sigma}_j\hat{\rho}\hat{\sigma}_i^\dagger
    -\frac{1}{2}
    \left\{
    \hat{\sigma}_i^\dagger\hat{\sigma}_j,
    \hat{\rho}
    \right\}
    \right),
    \label{eq:app:BM_reduced}
\end{align}
with the dissipator unchanged and the effective Hamiltonian now carrying both contributions,
\begin{equation}
    \frac{\hat{H}_{\mathrm{eff}}}{\hbar}
    =
    \sum_{i,j}\Delta^r_{ij}(\omega_0)\,\hat{\sigma}_i^\dagger\hat{\sigma}_j
    -
    \sum_{i,j}\Delta^{c}_{ij}(\omega_0)\,\hat{\sigma}_i\hat{\sigma}_j^\dagger.
    \label{eq:app:Heff}
\end{equation}
Using $\hat{\sigma}_i\hat{\sigma}_j^\dagger=\hat{\sigma}_j^\dagger\hat{\sigma}_i$ for $i\neq j$, while leaving the diagonal terms in their natural $\hat{\sigma}_i\hat{\sigma}_i^\dagger$ form, leads to the following form of the effective Hamiltonian:
\begin{equation}
    \frac{\hat{H}_{\mathrm{eff}}}{\hbar}
    =
    \sum_{i\neq j}\left[\Delta^{r}_{ij}(\omega_0)-\Delta^{c}_{ji}(\omega_0)\right]\hat{\sigma}_i^\dagger\hat{\sigma}_j
    +
    \sum_{i}\Delta^{r}_{ii}(\omega_0)\,\hat{\sigma}_i^\dagger\hat{\sigma}_i
    -
    \sum_i\Delta^{c}_{ii}(\omega_0)\,\hat{\sigma}_i\hat{\sigma}_i^\dagger,
    \label{eq:app:Heff_normal}
\end{equation}
where the diagonal resonant term $\Delta^{r}_{ii}\,\hat{\sigma}_i^\dagger\hat{\sigma}_i$ is the resonant part of the single-emitter Lamb shift, and the diagonal counter-rotating piece $-\sum_i\Delta^{c}_{ii}\,\hat{\sigma}_i\hat{\sigma}_i^\dagger$ arises from the anti-resonant (energy-nonconserving) process, acting only on the ground-state population. These terms can then either be absorbed into $\hat{H}_S$ or retained explicitly. Discarding the counter-rotating coupling from the outset ($\tilde{g}_{\alpha,\kk}\to0$, a premature rotating-wave approximation) sets $\Delta^{c}\to0$ and recovers the master equation of Eq.~\eqref{eq:BMa} exactly.

The difference of the two off-diagonal coefficients is the physical coherent coupling that enters the master equation of Eq.~\eqref{eq:BMa},
\begin{equation}
    J_{ij}(\omega_0)\equiv\Delta^{r}_{ij}(\omega_0)-\Delta^{c}_{ji}(\omega_0),
    \qquad(i\neq j).
    \label{eq:app:Jdef}
\end{equation}

We now write this coupling in compact form. Throughout, we invoke the relevant-band approximation $\omega_\alpha(\kk)\approx\omega_0$, justified by the off-resonant suppression of far-detuned bands, applying it to the non-singular frequency factors while keeping the resonant denominator $\omega_\alpha(\kk)-\omega_0$ intact. Using the time-reversal symmetry of the lossless dielectric [$\mathbf{f}_{\alpha,-\kk}(\rr)=\mathbf{f}^*_{\alpha,\kk}(\rr)$, $\omega_\alpha(-\kk)=\omega_\alpha(\kk)$] and the invariance of the mode sum under $\kk\to-\kk$, the counter-rotating shift can be expressed through the co-rotating couplings,
\begin{equation}
    \Delta^{c}_{ji}(\omega_0)
    =
    \sum_{\alpha,\kk}
    \frac{\tilde{g}_{\alpha,\kk}(\rr_j)\,\tilde{g}^*_{\alpha,\kk}(\rr_i)}{\omega_\alpha(\kk)+\omega_0}
    =
    \sum_{\alpha,\kk}
    \frac{g_{\alpha,\kk}(\rr_i)\,g_{\alpha,\kk}^*(\rr_j)}{\omega_\alpha(\kk)+\omega_0}.
    \label{eq:app:Deltac_TRS}
\end{equation}
Inserting Eqs.~\eqref{eq:app:JGammaBM} and \eqref{eq:app:Deltac_TRS} into Eq.~\eqref{eq:app:Jdef} and combining the two poles,
\begin{align}
    J_{ij}(\omega_0)-\ii\frac{\Gamma_{ij}(\omega_0)}{2}
    &=
    -\sum_{\alpha,\kk}
    g_{\alpha,\kk}(\rr_i)\,g_{\alpha,\kk}^*(\rr_j)
    \left[
    \frac{1}{\omega_\alpha(\kk)-\omega_0-\ii 0^+}
    +
    \frac{1}{\omega_\alpha(\kk)+\omega_0}
    \right]
    \nonumber\\
    &=
    -2\sum_{\alpha,\kk}
    \frac{g_{\alpha,\kk}(\rr_i)\,g_{\alpha,\kk}^*(\rr_j)\,\omega_\alpha(\kk)}
    {\omega_\alpha^2(\kk)-(\omega_0+\ii 0^+)^2}
    \nonumber\\
    &\underset{\omega_\alpha\approx\omega_0}{=}
    -2\omega_0\sum_{\alpha,\kk}
    \frac{g_{\alpha,\kk}(\rr_i)\,g_{\alpha,\kk}^*(\rr_j)}
    {\omega_\alpha^2(\kk)-(\omega_0+\ii 0^+)^2}.
    \label{eq:app:compact}
\end{align}

The relevant-band approximation enters in a single place, the numerator $2\omega_\alpha(\kk)\to2\omega_0$, while the resonant denominator is kept exact. Consequently, the imaginary part still returns the exact collective decay rate,
\begin{equation}
    \Gamma_{ij}(\omega_0)
    =
    2\pi\sum_{\alpha,\kk}
    g_{\alpha,\kk}(\rr_i)\,g_{\alpha,\kk}^*(\rr_j)\,
    \delta[\omega_\alpha(\kk)-\omega_0],
\end{equation}
whereas the anti-resonant pole, having no support on resonance, contributes only to the coherent coupling $J_{ij}$.

We now confirm that the same coefficient follows directly from macroscopic QED. Substituting the spectral representation of the photonic Green's tensor [Eq.~\eqref{eq:GFspectral1}, retaining both poles through the denominator $\omega_\alpha^2(\kk)-(\omega_0+\ii 0^+)^2$] into Eq.~\eqref{eq:coeffG}, and using
\begin{equation}
    \pp_i^*\cdot
    \left[
    \mathbf{f}_{\alpha,\kk}(\rr_i)\otimes
    \mathbf{f}_{\alpha,\kk}^*(\rr_j)
    \right]
    \cdot\pp_j
    =
    (\pp_i^*\!\cdot\mathbf{f}_{\alpha,\kk}(\rr_i))
    (\pp_j\!\cdot\mathbf{f}_{\alpha,\kk}^*(\rr_j)),
\end{equation}
yields
\begin{equation}
    J_{ij}(\omega_0)-\ii\frac{\Gamma_{ij}(\omega_0)}{2}
    =
    -\frac{\omega_0^2}{\hbar\varepsilon_0 Na^3}
    \sum_{\alpha,\kk}
    \frac{
    (\pp_i^*\!\cdot\mathbf{f}_{\alpha,\kk}(\rr_i))
    (\pp_j\!\cdot\mathbf{f}_{\alpha,\kk}^*(\rr_j))
    }{
    \omega_\alpha^2(\kk)-(\omega_0+\ii 0^+)^2
    }.
\end{equation}
With $C_{\alpha,\kk}^2=\hbar\omega_\alpha(\kk)/(2N a^3\varepsilon_0)$ [Eq.~\eqref{eq:app:Cnk}] and the definition of $g_{\alpha,\kk}(\rr_i)$ [Eq.~\eqref{eq:couplingeig}], the numerator equals $[2\hbar Na^3\varepsilon_0/\omega_\alpha(\kk)]\,g_{\alpha,\kk}(\rr_i)g_{\alpha,\kk}^*(\rr_j)$, so that
\begin{align}
    J_{ij}(\omega_0)-\ii\frac{\Gamma_{ij}(\omega_0)}{2}
    &=
    -2\omega_0^2
    \sum_{\alpha,\kk}
    \frac{g_{\alpha,\kk}(\rr_i)\,g_{\alpha,\kk}^*(\rr_j)}
    {\omega_\alpha(\kk)\left[\omega_\alpha^2(\kk)-(\omega_0+\ii 0^+)^2\right]}
    \nonumber\\
    &\underset{\omega_\alpha\approx\omega_0}{=}
    -2\omega_0
    \sum_{\alpha,\kk}
    \frac{g_{\alpha,\kk}(\rr_i)\,g_{\alpha,\kk}^*(\rr_j)}
    {\omega_\alpha^2(\kk)-(\omega_0+\ii 0^+)^2},
\end{align}
where the relevant-band approximation is applied to the prefactor, $\omega_0^2/\omega_\alpha(\kk)\to\omega_0$, again leaving the resonant denominator exact. This coincides with the reduced-model result of Eq.~\eqref{eq:app:compact}.

Both derivations therefore yield the same coherent coupling $J_{ij}$ and collective decay rate $\Gamma_{ij}$: the perturbative limit of the reduced microscopic model and the macroscopic-QED coefficient of Eq.~\eqref{eq:coeffG} are equivalent in the relevant-band regime.

\section{Numerical implementation of real-time dynamics and absorbing boundaries}~\label{app:boundaries}

The real-time simulations are performed on a finite $N\times N = 50\times 50$ square lattice obtained by truncating the infinite reduced photonic model. In a finite system, hard boundaries would reflect outgoing photonic wavepackets back into the bulk, where they would re-encounter and perturb the emitter. To suppress these finite-size artifacts, we introduce an absorbing boundary region of thickness $N_\mathrm{abs}=10$ unit cells along each edge, following a standard complex absorbing-potential strategy inspired by perfectly matched layers~\cite{berenger1994perfectly}.

Within the single-excitation sector, the full light-matter Hilbert space is spanned by states containing either one excited emitter or one photonic excitation in the reduced lattice basis. Projecting the full Hamiltonian onto this subspace yields a finite matrix representation
\begin{equation}
    \mathbf{H}=
    \begin{pmatrix}
        \mathbf{H}_S & \mathbf{H}_{\mathrm{int}} \\
        \mathbf{H}_{\mathrm{int}}^\dagger & \mathbf{H}_B
    \end{pmatrix},
    \label{eq:H_block}
\end{equation}
of dimension $N_\mathrm{tot} \times N_\mathrm{tot}$, where $N_\mathrm{tot} = N_e+N_oN$, $N_e$ is the number of emitters and $N_o$ the number of photonic orbitals per unit cell. The time evolution is then obtained by direct matrix exponentiation:
\begin{equation}
    \ket{\Psi(t)}=
    e^{-\ii \mathbf{H}t}
    \ket{\Psi(0)}.
\end{equation}

To implement absorption at the boundaries, we add site-dependent losses to the photonic sector through an imaginary onsite potential,
\begin{equation}
    \mathbf{H}_{B,*}
    =
    \mathbf{H}_B
    -
    \ii\,\mathrm{diag}
    (\kappa_1,\ldots,\kappa_{N_oN}),
    \label{eq:HB_pml}
\end{equation}
where $\kappa_{\alpha,\RR}\ge 0$ is the damping rate assigned to orbital $\alpha$ at lattice site $\RR$. The resulting effective Hamiltonian
\begin{equation}
    \mathbf{H}_*=
    \begin{pmatrix}
        \mathbf{H}_S & \mathbf{H}_{\mathrm{int}} \\
        \mathbf{H}_{\mathrm{int}}^\dagger & \mathbf{H}_{B,*}
    \end{pmatrix}
\end{equation}
is non-Hermitian, so the norm of
\begin{equation}
    \ket{\Psi_*(t)}=
    e^{-\ii\mathbf{H}_*t}
    \ket{\Psi(0)}
\end{equation}
is no longer conserved. This norm loss represents photon population irreversibly absorbed at the system boundaries.

The damping profile is chosen as a quadratic ramp,
\begin{equation}
    \kappa_{\alpha,\RR}=
    \begin{cases}
        \kappa_\mathrm{max}
        \left(
        \dfrac{N_\mathrm{abs}-d_\RR}{N_\mathrm{abs}}
        \right)^2,
        & d_\RR<N_\mathrm{abs},
        \\[6pt]
        0,
        & \text{otherwise},
    \end{cases}
    \label{eq:gamma_profile}
\end{equation}
where
\begin{equation}
    d_\RR=
    \min
    \left(
    \frac{N}{2}-|x_\RR|,
    \frac{N}{2}-|y_\RR|
    \right),
    \label{eq:dist}
\end{equation}
with $\RR=(x_\RR,y_\RR)$. Thus, the absorption increases smoothly from zero in the bulk to $\kappa_\mathrm{max}$ at the outer boundary, minimizing spurious reflections caused by abrupt impedance mismatches.

For the simulations shown in the main text, we use $N_\mathrm{abs}=10$ and $\kappa_\mathrm{max}=0.5\,c/a$. These values were selected from independent convergence tests monitoring the single-emitter decay dynamics while varying both parameters. The chosen values suppress boundary reflections while keeping the absorbing region sufficiently narrow that the bulk dynamics remain unaffected.

\section{Figures of merit used in the dynamical analysis}~\label{app:measures}

For the single-emitter spontaneous-emission problem, the total state remains in the single-excitation sector and can be written as
\begin{equation}
|\Psi(t)\rangle =
C_e(t)\,|e\rangle\otimes|\mathrm{vac}\rangle
+
\sum_{\RR} C_{\RR}(t)\,|g\rangle\otimes \hat{a}_{\RR}^\dagger|\mathrm{vac}\rangle.
\end{equation}
Tracing out the photonic degrees of freedom yields the reduced density matrix of the emitter:
\begin{equation}
\hat{\rho}_e(t)=
|C_e(t)|^2 |e\rangle\langle e|
+
\left(1-|C_e(t)|^2\right)|g\rangle\langle g|.
\end{equation}
The light-matter entanglement entropy is then quantified through the von Neumann entropy:
\begin{equation}
S(t)=
-\mathrm{Tr}\left[\hat{\rho}_e(t)\ln\hat{\rho}_e(t)\right],
\end{equation}
which directly yields Eq.~\eqref{eq:entanglement} of the main text.

To quantify the degree of directional emission, we partition the photonic crystal into the four quadrants centered at the emitter position and define the coarse-grained photonic populations:
\begin{equation}
P_q(t)=\sum_{\RR\in q}|C_{\RR}(t)|^2,
\end{equation}
where $q=1,\dots,4$ labels the four quadrants. The directional Shannon entropy is then defined as
\begin{equation}
S_{\mathrm{dir}}(t)=
-\sum_{q=1}^{4} P_q(t)\ln P_q(t).
\end{equation}
A value $S_{\mathrm{dir}}=\ln 4$ corresponds to uniform emission into all four quadrants, while $S_{\mathrm{dir}}=\ln 2$ indicates emission predominantly distributed between two opposite quadrants.

For the two-emitter dynamics discussed in Sec.~\ref{subsec:spontaneousemission}, the global state remains within the single-excitation manifold:
\begin{equation}
|\Psi(t)\rangle=
C_e(t)|eg\rangle\otimes|\mathrm{vac}\rangle
+
C_p(t)|ge\rangle\otimes|\mathrm{vac}\rangle
+
\sum_{\RR}C_{\RR}(t)|gg\rangle\otimes \hat{a}_{\RR}^\dagger|\mathrm{vac}\rangle.
\end{equation}

Tracing out the photonic bath, the reduced two-emitter density matrix is
\begin{equation}
\hat{\rho}_{ep}(t)
=
\mathrm{Tr}_{B}\left[|\Psi(t)\rangle\langle\Psi(t)|\right].
\end{equation}
Using the two-emitter basis
$\{|ee\rangle,|eg\rangle,|ge\rangle,|gg\rangle\}$, this density matrix takes the form
\begin{equation}
\hat{\rho}_{ep}(t)=
\begin{pmatrix}
0 & 0 & 0 & 0 \\
0 & |C_e(t)|^2 & C_e(t)C_p^*(t) & 0 \\
0 & C_p(t)C_e^*(t) & |C_p(t)|^2 & 0 \\
0 & 0 & 0 & \sum_{\RR}|C_{\RR}(t)|^2
\end{pmatrix}.
\end{equation}
The last diagonal element corresponds to the probability that both emitters are in their ground state while the excitation is in the photonic bath. For a two-qubit density matrix of this X-state form, the Wootters concurrence~\cite{garcia2023probing,gonzalez2015chiral,wootters1998entanglement} is
\begin{equation}
\mathcal{C}(t)
=
2\max\left[
0,
|\rho_{eg,ge}|
-
\sqrt{\rho_{ee,ee}\rho_{gg,gg}}
\right].
\end{equation}
Since $\rho_{ee,ee}=0$ in the single-excitation manifold and
$\rho_{eg,ge}=C_e(t)C_p^*(t)$, this reduces to
\begin{equation}
\mathcal{C}(t)=2|C_e(t)C_p(t)|.
\end{equation}

This quantity vanishes for separable emitter states and reaches unity for a maximally entangled Bell state.

\end{document}